\documentclass[12pt]{article}
\usepackage{epsfig}
\newcommand{\be}{\begin{equation}}
\newcommand{\ee}{\end{equation}}
\newcommand{\bea}{\begin{eqnarray}}
\newcommand{\eea}{\end{eqnarray}}
\newcommand{\ta}{\tilde\alpha}
\newcommand{\tb}{\tilde\beta}

\newcommand{\bn}{\bar\nu}

\newcommand{\LR}{$SU(3)\times SU(2)_L \times SU(2)_R\times U(1)_{B-L}\quad $}
\newcommand{\PS}{$ SU(4)\times SU(2)_L\times SU(2)_R\quad  $}

\newcommand{\at}{\tilde a}
\newcommand{\pt}{\tilde p}
\newcommand{\omt}{\tilde \omega}

\newcommand{\sgt}{\tilde \sigma}
\newcommand{\sgbt}{\tilde {\overline{ \sigma}}}

\newcommand{\sq}{\sqrt{2}}

\newcommand{\sqt}{\sqrt{3}}

\newcommand{\s}{\Sigma}
\newcommand{\Sigb}{{\overline\Sigma}}
\newcommand{\oot}{\overline {126}}

\newcommand{\nnu}{\nonumber\\}
\def\blfootnote{\xdef\@thefnmark{}\@footnotetext}
\begin{document}
\begin{titlepage}
\vspace{4\baselineskip}
\begin{center}{\Large\bf  MSGUT : From Bloom to Doom
 }

\end{center}
\vspace{1cm}
\begin{center}
{\large
 Charanjit {Singh Aulakh}
 \footnote{E-Mail:  aulakh@pu.ac.in   }
   and Sumit Kumar Garg
 }
\end{center}
\vspace{0.2cm}
\begin{center}
  {\it
Dept. of Physics, Panjab University,\\ Chandigarh, India 160014}
\end{center}
\vspace{1cm}
\begin{abstract}

By a systematic  survey of the parameter space we confirm our
surmise\cite{gmblm} that the Minimal Supersymmetric GUT(MSGUT)
based on the ${\bf{210\oplus   126\oplus {\overline {126}}\oplus
10 }}$ Higgs system is incompatible with the generic Type I and
Type II seesaw mechanisms.   The incompatibility of the Type II
seesaw mechanism with this MSGUT is due to its generic extreme
sub-dominance with respect to the Type I contribution. The Type I
mechanism although dominant over Type II is itself unable to
provide Neutrino masses  larger than  $ \sim 10^{-3}$  eV anywhere
in the parameter space. Our Renormalization Group  based analysis
shows the origin of these difficulties to lie in a conflict
between baryon stability and neutrino oscillation. The MSGUT
completed with a {\bf{120}}-plet Higgs is the natural next to
minimal candidate. We propose a scenario where the {\bf{120}}-plet
collaborates with the {\bf{10}}-plet to fit the charged fermion
masses. The freed {{\bf{126}}}-plet couplings can then  give
sub-dominant contributions to charged fermion masses
{\textit{and}} enhance the Type I seesaw masses sufficiently to
provide a  viable seesaw mechanism. We give formulae required to
verify this scenario.

\end{abstract}
\end{titlepage}

\normalsize\baselineskip=15pt

\section{ Introduction}

 SO(10)  GUTs   accomodate complete fermion families together
 with the superheavy right
handed neutrinos (required by the seesaw
mechanism\cite{seesaw,seesaw2}) in a   {\bf{16}}-plet spinorial
irreps. They also provide very natural Higgs multiplets ($\bf{{
\oot}})$ (or $\bf{\overline{16}_H}\times \bf{\overline{16}_H} $)
to generate the ($U(1)_{B-L}$ breaking) right handed neutrino
Majorana masses and the   ($SU(2)_L \times U(1) $ breaking)
triplet vevs required to implement the Type I\cite{seesaw} and
Type II\cite{seesaw2} mechanisms respectively. Thus models which
incorporate  high scale breaking of the $SU(2)_R \times
U(1)_{B-L}$  components of the SO(10) gauge symmetry provide an
elegant and natural context to understand the clear indication of
a  seesaw connection between neutrino mass and GUT mass scales
provided by the discovery of neutrino masses in the 50 milli-eV
range  ($\Delta m_{\nu}^2 \sim 2.2 \times 10^{-3} eV^2\sim ((10^2
GeV)^2/(10^{15} GeV))^2 $ ). In supersymmetric models with left
right symmetric (\LR$ \subset G$) gauge groups, R-parity
($R_p=(-)^{3(B-L)+2S}$)\cite{rparityBL} is a part of the Gauge
symmetry and is naturally
preserved\cite{abs,ams,amrs98,genea,abmrs01,abmsv} till low
energies in models using only B-L even  vevs for the seesaw. They
thus predict a stable LSP  which is  welcome as cosmological dark
matter. With such a complement of virtues Supersymmetric SO(10)
GUTs of some variety or the other are considered leading
contenders by a majority of workers. The Supersymmetric SO(10) GUT
based on the ${\bf 126} \oplus {\bf {\overline{126}}} \oplus  {\bf
210}$ Higgs multiplets proposed long ago \cite{aulmoh,ckn}  shares
all these manifest virtues of LR supersymmetric models.  It  has
the additional virtue of maximal simplicity from a parameter
counting and representation economy point of
view\cite{aulmoh,ckn,babmoh,abmsv}. It thus lays claim to the role
of contributing the correct GUT symmetry breaking sector (AM
Higgs) for a  fully realistic Minimal Supersymmetric GUT (MSGUT).
The other crucial component  required to fully define a MSGUT is
however the complement of fermion mass (FM)Higgs used to fit the
charged fermion masses and mixings {\it{and}} the neutrino
oscillation mass and mixing parameters which constitute the actual
`data' that a GUT must confront. SO(10) permits  an FM Higgs
system based on $ {\bf {10}} \oplus{\bf {120}} \oplus{\bf
{\overline{126}}} $ irreps. The first and third of these  are have
Yukawa couplings to the matter  \textbf{16}-plets   which are
symmetric under interchange of family indices   and have been
extensively considered.  The $\bf{120}$-plet with its
antisymmetric Yukawa couplings has so far been conceded  much less
 importance, acting either as a perturbation to the main structure
 determined by the $\mathbf{10} $ and $ \mathbf{\overline{126}} $
 plets \cite{moh120,bert} or else somewhat cursorily
 \cite{matsuda,oshimo,bs,bmsv3}. In this paper we shall develop a
 line of argument that culminates in a suggestion  that a  central role for
 the $\mathbf{120}$-plet is natural.

 In 1992, with LEP data in
hand,   Babu and Mohapatra\cite{babmoh}  proposed that if only the
 ${\bf{10\oplus \oot}}$ irreps were used to fit the
 charged fermion masses and mixing  then the matter fermion
 Yukawa  couplings  of the GUT would
 be completely determined.  Hence the model would become
 predictive in the   neutrino sector. The  compatibility of this
 scenario with the (now better known) neutrino mass-mixing  data
  has been the subject of an accelerating succession of works
  since their proposal   \cite{japsnu,matsu0,bsv,gohmoh,bert,babmacesnu}.
  These works have demonstrated the compatibility of the data with
  the  generic (Georgi-Jarlskog \cite{georjarls}
 plus Type I and Type II seesaw)  structural form of the fermion mass formulae
  dictated by SO(10) clebsches and restricted FM Higgs content.
   Symmetric Yukawa couplings alone are quite successful
   but the possibility  of  improving  the fits, particularly as regards the CKM CP phase,
   by  using a  subdominant contribution of    the {\bf{120}}-plet
   has also been demonstrated \cite{bert,moh120}.  However the generic
   fitting procedure  does not fix the over all mass scale of the
   neutrino masses or the relative strength of the Type I and Type
   II masses. On the other hand these quantities {\it{are}} fixed
   in a fully specified GUT model and it is thus crucial to
   investigate whether any given GUT model of this general
   type is compatible with both  the generic fits and the actual
   neutrino mass data.
In the  1990's   the construction of natural and  fully consistent
Minimal Left Right Supersymmetric models  (MSLRMs) was
accomplished\cite{abs,ams,amrs98}. Moreover these analyses
  showed  clearly\cite{abmrs99,abmrs01}  what was already noted at
 the beginning\cite{aulmoh,aulphd} of the study of
 multi-scale Susy  GUTs : that    in LR Susy GUTs there
  are light multiplets which
  violate  the conventional wisdom   of the ``survival
  principle''.  Thus it was clear\cite{abmrs01,abmrs99}
  that Susy GUTs  required the use of {\it{calculated}}
   rather than merely  (``survival principle'')
   estimated masses for RG analyses.
The RG analysis carried out in this work already indicated that
the use  of calculated spectra   forces together the various
 possible intermediate  scales into a narrow range close
 to the GUT scale   resulting in an effective  ``SU(5) conspiracy''
   i.e the necessity for single step breaking of SO(10).

  A  calculation of the   full GUT spectrum and couplings in various SO(10)
  GUT models required complete knowledge of the "Clebsches"  relating  $
  SO(10)$ and $ G_{SM}\times U(1)_R $  group
  labels on SO(10) fields - at least for tensorial
irreps -   before progress in the    Renormalization Group (RG)
analysis  could  be made. Furthermore the compatibility of
  fermion fits with GUTs could also be investigated only if
 Clebsches were also available for the spinorial
 (${\bf{16}}$-plet)  irrep.  The requisite computations
 (based on   decomposition of SO(10)
labels into those of the \PS  (``Pati-Salam'') sub-group ) were
 presented  by us in \cite{ag1}. Using these Clebsches first
 the mass matrix of the MSSM type doublets ($[1,2,\pm 1]$)
 and classic GUT $\Delta B\neq 0$ process mediating
  triplets ( $[3,1,\pm {2\over 3}]$ \cite{ag1} and then
 a complete calculation\cite{ag2}, of {\it{all}} the
 couplings and mass matrices of the MSGUT was given by us.
   With similar motivations two calculations, one in parallel
 and cross checking with ours\cite{bmsv}, and another
 \cite{fuku04} quite separate, which  both used the same  somewhat
 abstract\cite{heme}  method (but different phase conventions)  to calculate the ``Clebsches''
 of  tensor (but not   spinor) multiplets, appeared.
 In \cite{ag1,ag2} we     provided  the
 complete (chiral and gauge)  spectra, neutrino mass matrices,
 gauge and chiral couplings and the
  effective $d=5$  operators for Baryon violation in terms
 of GUT parameters and MSSM fields. Thus    the  stage was laid both  for a
 completely explicit RG based analysis of the MSGUT
 and a completely specified investigation of the
 compatibility of Type I and Type II seesaw mechanisms.
      The  calculation of threshold effects based on these
 spectra  was described in\cite{ag2,gmblm} and shown
  to controvert the conventional wisdom\cite{dixitsher} which doubted the stability
  of the successful   unification   (and $ m_t,\sin^{2} \theta_{W} $
  prediction\cite{marsenj})  of the one-loop MSSM coupling
  evolution\cite{marsenj,jones,amaldi}.

 The question of the fitting of the charged fermion
  masses, the relative size of the neutrino mass splittings
   and the quark and lepton mixing matrices had been
  extensively analyzed\cite{babmoh,japsnu,matsuda,bsv,gohmoh,bert,babmacesnu}
   using the generic formulae valid in SO(10)
  GUTs with only\cite{babmoh} ${\bf{10,\oot}}$ FM Higgs
  representations. However this procedure does
  not constrain the relative size of the Type I and Type II
  contributions  or the overall   scale
  of the neutrino masses.  Within a fully specified theory like the
  MSGUT, however,  both these parameters are specified. In
  \cite{gohmohnasri} the question of assuring Type II (over Type I) dominance on
  the basis of the Yukawa values found in the  successful generic Type II
  fits\cite{bsv,gohmoh} was considered. The authors concluded on the basis of
  order of magnitude estimates that
  the Type I mass would emerge as {\it{too large }} unless the
  $B-L$ breaking scale was raised. Furthermore they found that
   the mass of the $SU(2)_L$   triplet field whose
   tadpole  vev gives rise to the Type II seesaw must be lowered
   to enhance Type II seesaw masses. To ensure these they
   considered extending the model by modifying the GUT scale
   breaking to be via $SU(5) $(this required introducing a $\bf{54}-plet$ )
   or else by perturbing it by introducing a ${\bf{120}}$-plet.

   Our point of view, program  and results on these questions are related but still
    quite distinct from the estimates and scenarios of \cite{gohmohnasri}.
   In\cite{gmblm}  we took up the question of the
  compatibility of the MSGUT with the successful generic  {\bf{10}} +
  ${\bf{\overline{126}}}$   FM Higgs fits.  We surveyed the
  variation of the relative strength of Type I and Type II seesaw
  mechanisms, as well as their absolute magnitudes while keeping
  in view the viability of the relevant regions of the GUT
  parameter space  i.e  with the stability of the basic unification
  scenario.
  Our  graphical survey   yielded the striking observations that :
\begin{itemize}
 \item {{\it For generic  as well as  special values of MSGUT
  couplings and FM Yukawas taken from the generic Type II  seesaw
  fit, the Type II seeaw itself implies that it   is highly subdominant to Type I seesaw
  in the MSGUT unless the relative strength of Type I versus Type II is adjusted
  to a very small value. The primary reason for this  relative  dominance of Type I
   is not the  value of the symmetry breaking  scale but rather
  the value of the effective coupling in the Type I seesaw
   formula implied by the Yukawas found in the Type II
   generic fit (which simply assumes Type II is dominant)}}

 \item  {\it{The maximal values of the Type I seesaw
  masses attainable in the MSGUT are at least one order of
  magnitude short of those required by atmospheric neutrino
  oscillations }}.

  \end{itemize}

  Note that on both accounts  our conclusions are at variance with
  the  \cite{gohmohnasri} although they also find difficulties in
  ensuring Type II dominance over Type I.

In the present paper we present a detailed survey of the parameter
space to definitely confirm the above conclusions and thus pose a
stringent challenge to the viability of the original proposal of
Babu and Mohapatra\cite{babmoh} in the context of the MSGUT.
Already in \cite{ag2} we noted that  the variation of the
unification stability monitoring parameters (USMPs)(i.e $\Delta
(Log M_X), \Delta (Sin^2\theta_W),\Delta(\alpha_G^{-1})$) with the
fast control parameter $\xi$ of the MSGUT   exhibited very
striking sharp peaks and dips (with the characteristic appearance
of poles of a function of $\xi$). In many cases these spikes were
at the values of points of increased symmetry uncovered by the
analytic solution of the AM spontaneous symmetry breaking problem
in the MSGUT\cite{abmsv,ag2,bmsv}. In addition there were other
spikes (not obviously related to points of extended symmetry)
whose investigation was called for by the ``tomogram''  of the
MSGUT parameter space that we had provided\cite{ag2}.
In\cite{abmsv,ag2,bmsv} an analytic parametrization of the SSB in
the MSGUT in terms of the solutions of a cubic equation(linear in
a fast control parameter $\xi$) was achieved. In particular in
\cite{abmsv,bmsv,bmsv2} the parameter $\xi$ was eliminated -using
the cubic  equation- for the solution values $x $ and thus a very
direct,elegant and unified parametrization  of the pole and zero
structures characterizing the MSGUT AM-SSB was achieved in terms
of the variation of a single complex variable x. This allowed the
identification of additional special points. In particular,
motivated by  our observations\cite{gmblm}, of a set of points
associated with possible growth of the Type I and Type II seesaw
contributions\cite{abmsv,bmsv,bmsv2}. Such growth had already been
previously observed at some points in \cite{gmblm} but found to
give unviable USMPs.
  Thus to complete the program of\cite{gmblm} we have now investigated
  the behaviour   of the Type I and Type II coefficient functions
  defined by the precise  clebsches and fermion mass matrices
   of the MSGUT\cite{ag2,gmblm} over   the complex  $x$ (  equivalently to $\xi$ ) plane :
    including all the special points described above.
      The loopholes offered\cite{gmblm,bmsv2} by the
   special points are illusory rather than real : typically some
   or all of the USMPs explode due to the very growth which is
   being invoked to strengthen one or the other seesaw.  Our
   conclusions are thus unchanged from those of
   \cite{gmblm}:  {\emph{\textbf{ the MSGUT has suffered a failure of its
   most critical features and is thus now defunct.}}}\\

A way out of this impasse  appears by allowing in the
{\bf{120}}-plet which was earlier discriminated against on grounds
of convenience and simplicity alone. `Lamppost logic' may have
failed here as so often before. Nature does not heed the wishful
thinking behind such logic: the morning often reveals that the
keys sought for all  night  lay in the gutter just beyond the
lamppost ! Another alternative is that SSB in the MSGUT may
actually be controlled \cite{trmin,tas} by the explosive growth of
the gauge coupling {\it{above}} $M_X$ due to gaugino condensation
in the coset $SO(10)/G_{123}$  which drives\cite{tas} chiral
condensates of the AM Higgs fields  whose value may be
{\it{calculable}} due to the power of supersymmetric holomorphy.
In such a scenario the crucial composition of the zero mode Higgs
doublets must  be evaluated without using the ``cubic'' solution
mentioned above.

   In Section \textbf{2}, referring the reader to the original papers
   \cite{ag1,ag2,gmblm} for all derivations,
    we  present the basic formulae\cite{ag2,gmblm}
    governing the Type I and II  seesaw mechanisms in the MSGUT
    and identify the  Seesaw Monitoring Parameters (SMPs)
    ($R,F_I,F_{II}$) that we will use for our graphical
    presentation of the  seesaw relevant
     topography of the parameter space.
     We  list  the obvious  points of enhanced gauge symmetry.
     This is  accompanied by  an
Appendix   where these formulae are also evaluated in the
(Rational) function (of $x$) form  advocated by\cite{bmsv2}. This
allows the identification of an additional list of exceptional
points requiring special treatment.
 In Section \textbf{ 3}  we  describe the analytic and  graphical investigation of the
 behaviour of the USMPs and SMPs as we scan over the complex $x$
 plane for generic and for exceptionally  favourable values of the `slow' AM Higgs
 parameters \cite{ag2} i.e. $x$ (equivalently $\xi$) .
   By covering the behaviour at the enhanced symmetry points and
   exceptional seesaw points
       (collectively called ESPs)and scanning the whole $x$-plane we
       eliminate all escape routes and pin down the MSGUT to its doom.
  In Section  \textbf{4} we propose that the inclusion of the {\bf{120}}-plet
  in a very specific role can solve the problem with the small neutrino masses
  found for  the MSGUT and give the fermion mass
    formulae for the NMSGUT along with the doublet Higgs mass
    matrix ${\cal{H}}$. This specifies
     the composition of the massless doublets once $det {\cal{H}}=0$ is
     imposed. Thus the stage is set for the accomplishment of  a
     program analogous to that accomplished  for
     the MSGUT
\cite{aulmoh,ckn,ag1,abmsv,ag2,bmsv,fuku04,fukrebut,vietproc,gmblm,bmsv3},
 which  can be implemented without ambiguity. We conclude with a
discussion of the outlook for future work.

\section{Seesaw formulae and SMPs for the MSGUT }
\subsection{MSGUT couplings, vevs  and masses }
 The MSGUT is the renormalizable  globally supersymmetric $SO(10)$ GUT
 whose Higgs chiral supermultiplets  consist of
  \textbf{A}djoint \textbf{M}ultiplet (\textbf{AM}) type   totally
 antisymmetric tensors : ${\bf{210}}(\Phi_{ijkl}),
 {\bf{\overline{126}}}({\bf{\Sigb}}_{ijklm}),$
 ${\bf{126}} ({\bf\Sigma}_{ijklm})(i,j=1...10)$ which   break the GUT symmetry
 to the MSSM, together with Fermion mass (FM) Higgs {\bf{10}}-plet(${\bf{H}}_i$).
  The  ${\bf{\overline{126}}}$ plays a dual or AM-FM
role since  it also enables the generation of realistic charged
fermion   and    neutrino masses and mixings (via the Type I
and/or Type II mechanisms);  three  {\bf{16}}-plets
${\bf{\Psi}_A}(A=1,2,3)$  contain the matter  including the three
conjugate neutrinos (${\bar\nu_L^A}$).

 The   superpotential   (see\cite{abmsv,ag1,ag2,bmsv} for
 comprehensive details ) contains the  mass parameters
 \bea
 m: {\bf{210}}^{\bf{2}} \qquad ; \qquad M : {\bf{126\cdot{\overline {126}}}}
 ;\qquad M_H : {\bf{10}}^{\bf{2}}
\eea and trilinear couplings
  \bea
 \lambda : {\bf{210}}^{\bf{3}} \qquad ; \qquad  \eta  :
 {\bf{210\cdot 126\cdot{\overline {126}}}}
 ;\qquad  \gamma \oplus {\bar\gamma}  : {\bf{10 \cdot 210}\cdot(126 \oplus
{\overline {126}}})
  \eea

The GUT scale vevs that break the gauge symmetry down to the SM
symmetry (in the notation of\cite{ag1})  are{\cite{aulmoh,ckn}}
\bea
 {\langle(15,1,1)\rangle}_{210}& :&
\langle{\phi_{abcd}}\rangle={a\over{2}}
\epsilon_{abcdef}\epsilon_{ef}\\
\hfil\break
\langle(15,1,3)\rangle_{210}~&:&~\langle\phi_{ab\ta\tb}\rangle=\omega
\epsilon_{ab}\epsilon_{\ta\tb}\quad\\
\quad\langle(1,1,1)\rangle_{210}~&:& ~\langle\phi_{ {\tilde
\alpha}{\tilde \beta} {\tilde \gamma}{\tilde \delta}}
\rangle=p\epsilon_{{\tilde \alpha} {\tilde \beta} {\tilde
\gamma}{\tilde \delta}}\quad\\
 \hfil\break
\langle(10,1,3)\rangle_{\oot} ~&:&
  \langle{\overline\Sigma}_{\hat{1}\hat{3}\hat{5}
\hat{8}\hat{0}}\rangle= \bar\sigma \quad\\
\langle({\overline{10}},1,3)\rangle_{126} ~&:&
\langle{\Sigma}_{\hat{2}\hat{4}\hat{6}\hat{7}\hat{9}}
\rangle=\sigma. \eea
 The vanishing of the D-terms of the SO(10) gauge sector
 potential imposes only the condition $
 |\sigma|=|{\overline{\sigma}}| $.
Except for the simpler cases corresponding to enhanced unbroken
gauge symmetry  ($SU(5)\times U(1), SU(5), G_{3,2,2,B-L},
G_{3,2,R,B-L}$ etc)\cite{abmsv,bmsv}, this system of equations is
essentially cubic and can be reduced to  the single equation
\cite{abmsv}
 for a variable $x= -\lambda\omega/m$, in terms of
 which the vevs $a,\omega,p,\sigma,
 {\overline\sigma}$ are specified  :
\be 8 x^3 - 15 x^2 + 14 x -3 = -\xi (1-x)^2 \label{cubic} \ee
where  $\xi ={{ \lambda M}\over {\eta m}} $. Then the
dimensionless vevs in units of (m/$\lambda$) are $\omt=-x$
\cite{abmsv} and \be \at={{ (x^2 +2 x -1)}\over (1-x)}\quad ;\quad
 \pt={{x(5 x^2-1)}\over {(1-x)^2}}\quad ; \quad
\sgt\sgbt={2\over \eta}{{\lambda x(1-3x)(1+x^2)}\over {(1-x)^2}}
\label{dlvevs}\ee
 This   exhibits the crucial importance of the
parameters $\xi,x$. Note that one can trade\cite{abmsv,bmsv,bmsv2}
the parameter $\xi$ for $x$ with advantage (using   equation
(\ref{cubic})) since $\xi$ is {\it{uniquely}} fixed  given  $x$.
By a survey of  the behaviour of the theory as a function of the
complex parameter $x$ we are simultaneously covering the behaviour
of the three different solutions possible for each complex value
of $\xi$.  We thus change our  graphical presentation in terms of
three plots versus $\xi$\cite{ag2,gmblm}  to a single plot versus
x for each USMP or SMP. Moreover, as emphasized by\cite{bmsv2}
analysis of the spikes observed in \cite{ag2,gmblm} is also
facilitated by the use of the parameter $x$.  Using the above vevs
and the methods of \cite{ag1} we calculated the complete gauge and
chiral multiplet GUT scale spectra {\it{and}} couplings for the 52
different MSSM multiplet sets falling into 26 different MSSM
multiplet types (prompting a natural alphabetization of their
naming convention\cite{ag1} !) of which 18 are unmixed while the
other 8 types occur in multiple copies.
   The (full details of these)  spectra may be found in\cite{ag1,ag2}
   and equivalent results(with slightly differing conventions)
    are presented in\cite{bmsv}. A related
   calculation with very different conventions has been reported
   in \cite{fuku04,fuku0405}. The initially controversial
   relation between the overlapping parts of these papers was
   discussed and resolved in \cite{fukrebut}.

Among the mass matrices   is the all important $4\times 4$ Higgs
doublet mass matrix \cite{ag1,ag2}  ${\cal H}$ : {{\scriptsize\bea
{\cal{H}}=\left({\begin{array}{cccc} -M_{H} &
+\overline{\gamma}\sqrt{3}(\omega-a) & -{\gamma}\sqrt{3}(\omega+a)
& -{\bar{\gamma}{\bar{\sigma}} }\\
 -\overline{\gamma}\sqrt{3}(\omega+a) & 0 & -(2M+4\eta(a+\omega)) &
0\\
\gamma\sqrt{3}(\omega-a) & -(2M+4\eta(a-\omega)) & 0 &
-{2\eta\overline{\sigma}\sqrt{3}}\\
  -{\sigma\gamma } &
-{2\eta\sigma\sqrt{3}} & 0 & {-{2m}+6\lambda(\omega-a)}
\end{array}}\right)  \eea}}
 ${\cal H}$  can be diagonalized
by a bi-unitary transformation\cite{abmsv,bmsv,ag2}: from the 4
pairs of Higgs doublets $h^{(i)},{\bar h}^{(i)}$ arising from the
SO(10) fields to a new set $H^{(i)},{\bar H}^{(i)}$ of fields in
terms of which the doublet mass terms  are diagonal. \bea
{\overline U}^T {\cal H}U &=&   Diag ( m_H^{(1)},m_H^{(2)},....)
 \nnu
 h^{(i)} &=& U_{ij} H^{(j)}  \qquad ;\qquad   {\bar h}^{(i)} = {\bar
U}_{ij} {\bar H}^{(j)}  \eea

 To keep one pair
of these doublets light one  tunes $M_H$ so that $Det{\cal H}=0$.
This matrix can then be  diagonalized  by a bi-unitary
transformation yielding thereby the coefficients describing the
proportion of the doublet fields in the ${\bf{10,\oot,126,210}}$
GUT multiplets  present in the light doublets : which proportions
are important for many phenomena.
 In the effective theory at low energies the
GUT Higgs doublets  $h^{(i)},{\bar h}^{(i)}$  are present in the
massless doublets   $H^{(1)},{\bar H}^{(1)}$ in a proportion
determined by the  first columns of the matrices  $U,{\bar U}$ :
    \bea E < <M_X \qquad : \qquad  {  h}^{(i)} &\rightarrow&  {  \alpha}_i {  H}^{(1)} \quad
  ; \quad
{ \alpha}_i = {  U}_{i1} \nnu
   {\bar h}^{(i)} &\rightarrow&  {\bar \alpha}_i {\bar H}^{(1)} \quad
  ; \quad
{\bar\alpha}_i = {\bar U}_{i1}
  \eea
The all important normalized 4-tuples $\alpha,\bar\alpha$ can be
easily determined by solving the zero mode conditions: ${\cal {H}}
\alpha =0 ~ ;~  \bar\alpha^T {\cal{H} }=0 $.

\subsection{ RG Analysis and   USMPs } In
\cite{ag2,gmblm} we discussed at length the use of plots of the
USMPs versus the fast parameter $\xi$ to investigate the
question :\\
 {\it{ Are the one loop values of $\alpha_G(M_X), Sin^2\theta_W $ and
 $M_X$ generically stable against superheavy threshold
corrections ?}}.\\
   We followed the approach of Hall\cite{hall} in which the
    mass
 ($ M_X = \frac{m g}{\lambda} \sqrt{ 4 |\tilde{a} + \tilde{w}|^2 +2 |\tilde{p}
     + \tilde{\omega}|^2}$)   of the   baryon number violating
superheavy  ($[3,2,\pm{5\over 3}]$ or  X-type) gauge bosons is
chosen  as the transition scale between the effective  MSSM  and
the full SO(10) GUT with all superheavy fields retained. Thus
$M_X$  is used as  common  \emph{physical}  superheavy matching
point ($M_i=M_V=M_X$) in the equations relating
  the   MSSM couplings to the SO(10)   coupling   :
\bea {1\over{\alpha_i(M_S)}}
 ={1\over{\alpha_G(M_X)}} +
 8 \pi b_i ln{{M_X}\over{M_S} }
  + 4 \pi \sum_j {{{b_{ij}} \over {b_j}}} ln X_j -
4\pi\lambda_i(M_X) \eea See \cite{hall,ag2} for details.   In this
approach it is recognized  that
 above the  scale $M_X$ the effective theory\emph{ changes} from the MSSM to
    Susy  $SO(10)$ model {\it{structured by the
complex superheavy spectra which we have computed }} so that the
   theory  appears as unbroken SO(10) only to leading order.
   Thus we compute the corrections to  the three USMPs
$Log_{10}{M_X}, sin^2\theta_W (M_S) , \alpha_G^{-1}(M_X) $
  as a function of the MSGUT  parameters  and the
  answer to the question of stability of the
 perturbative unification is determined by  the
 ranges of GUT parameters where these corrections are
 consistent with the known or surmised data on
 $Log_{10}{M_X}, sin^2\theta_W (M_S),\alpha_G $.   The consistency
  requirement that the   SO(10) theory remain perturbative
  after threshold and two loop correction
   and $\alpha_G$ not decrease so much as to   invalidate the neglect
  of one-loop effects in the couplings
  $ \lambda,\eta,\gamma,\bar{\gamma}$ is also appropriate.
  We  implement these  by the requirements  that the
  USMPs   obey
\bea
|\Delta_G|&= & |\Delta  (\alpha_G^{-1}(M_X))| \leq 10 \nonumber \\
\Delta_X &=&\Delta (Log_{10}{M_X}) \geq -1 \nonumber \\
\Delta_{W}&=&  |\Delta(sin^2\theta_W (M_S))|< .02
\label{USMPlimits} \eea See \cite{ag2,gmblm} for a detailed
discussion.
  We find the corrections
\bea
 \Delta^{(th)}(Log_{10}{M_X})  &=& .0217 +.0167 (5 {{\bar b}'}_1
 +3{{\bar b}'}_2 -8
  {{\bar b}'}_3) Log_{10}{{M'}\over  {M_X }} \label{Deltasw}\nnu
\Delta^{(th)} (sin^2\theta_W (M_S)) &=&
   .00004 -.00024 (4 {{\bar b}'}_1 -9.6 {{\bar b}'}_2 +5.6
  {{\bar b}'}_3) Log_{10}{{M'}\over  {M_X }}
  \label{Deltath}\nnu
\Delta^{(th)} (\alpha_G^{-1}(M_X)) &=& .1565 +  .01832 (5 {{\bar
b}'}_1 + 3 {{\bar b}'}_2 + 12 {{\bar b}'}_3) Log_{10}{{M'}\over
{M_X }}
  \eea
Where ${\bar b'}_i = 16\pi^2 b_i'  $ are   1-loop $\beta$ function
coefficients ($ \beta_i=b_i g_i^3 $)
 for multiplets with  mass $M'$ (a sum over representations
  is implicit).
 Note a minor, but crucial correction,  in the above equations
 relative to \cite{ag2,gmblm}: there and here,
 we actually used $M'\over M_X$ as
 the argument of the logarithms in these formulae. This is
 the value given by the algebra, which is then  conventionally
 approximated by $M'\over M_X^0$ since the difference is   of
order $g$\cite{hall}. However in our case we can retain the
``exact" form and still calculate  the threshold corrections since
those depend on ratios of masses. This improves the analysis
\emph{and} makes it more convenient. Moreover (see below) it means
that the overall scale parameter that we use , namely the
coefficient $m$ of $ \textbf{210}^2$ in the superpotential, is not
fixed at the one loop unification scale $10^{16.25} GeV$ but
slides with the corrections to $M_X$. These corrections are to be
added to the one loop values corresponding to the successful gauge
unification of the MSSM  : Using the values
  \bea
    \alpha_G^0(M_X)^{-1}= 25.6\quad ;
    \quad M_X^0=10^{16.25} ~GeV \quad ;\quad
 M_S=1 ~  TeV \nonumber \\
 \alpha_1^{-1}(M_S)=57.45 \quad ;\quad
  \alpha_2^{-1}(M_S)=30.8 \quad ;
  \quad\alpha_3^{-1}(M_S)=11.04
 \eea
 the two loop corrections are
\bea
 \Delta^{2-loop}(log_{10}{{M_X}\over {M_S}})&=&-.08 \qquad ;
  \quad \Delta^{2-loop}(sin^2\theta_W (M_S))= .0026
\nnu
 \Delta^{2-loop}\alpha_G^{-1}(M_X)&=& -.546
  \eea
We see that in comparison with the large threshold effects that
might be  expected\cite{dixitsher}  in view of the large number
(504) of heavy fields    the 2 loop corrections   are quite
small\cite{ag2}. The striking new result of \cite{ag2} was thus
  the explicit demonstration that the
conjecture\cite{dixitsher} concerning the ``Futility of high
precision SO(10) calculations'' was in fact false and that a
proper calculation based on calculated  spectra yields well
defined and consistent results for significant regions
of the MSGUT parameter space (see below).   \\
  The parameter $\xi= \lambda M/ \eta m$ is the only
  numerical parameter that  enters into the cubic
   eqn.(\ref{cubic})  that determines the parameter $x$
   in terms of which all the  superheavy vevs are given.
    {\it{ It is thus  the most crucial  determinant of
     the mass spectrum }}.  The dependence
of the threshold corrections on the parameters
 ${\lambda,\eta,\gamma,{\bar\gamma}}$  is comparatively mild  except
when coherent e.g when  many masses are lowered  together leading
to $\alpha_G $ explosion, $ Log M_X$ collapse or large changes
 in $sin^2\theta_W (M_S)$.

The typical behaviours of the USMPs can be seen in Figs 1-6.
\begin{figure}[h!]
\begin{center}
\epsfxsize15cm\epsffile{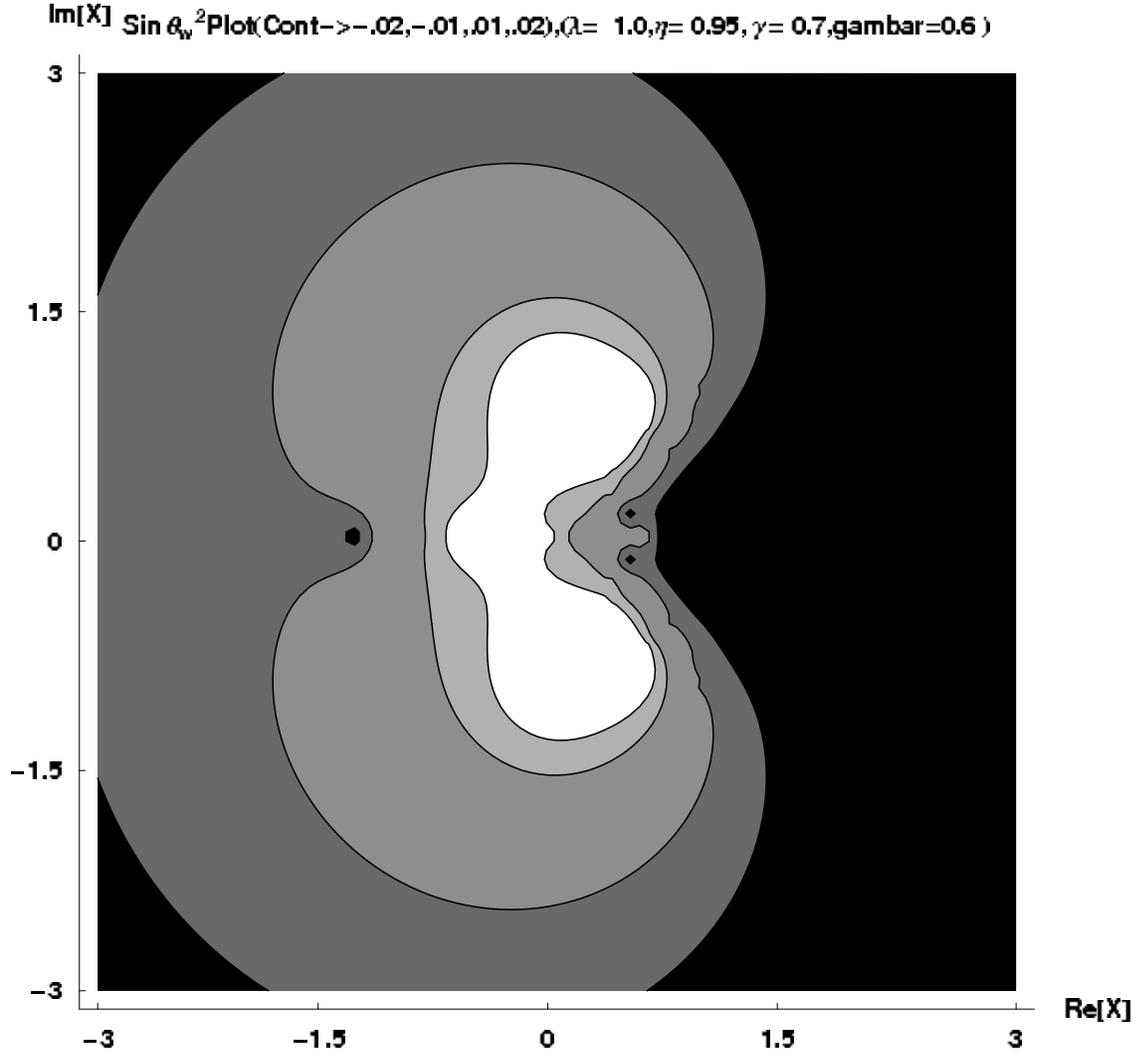}
 \caption{  Contour Plot of the threshold
corrections to $Sin^2\theta_w$  on the complex $x$ plane. Contours
at $\Delta_W =-.02,-.01,.01,.02$.  The shading progresses from
black ($<-.02$) to white $>.02$. }
\end{center}
\end{figure}

\begin{figure}[h!]
\begin{center}
\epsfxsize15cm\epsffile{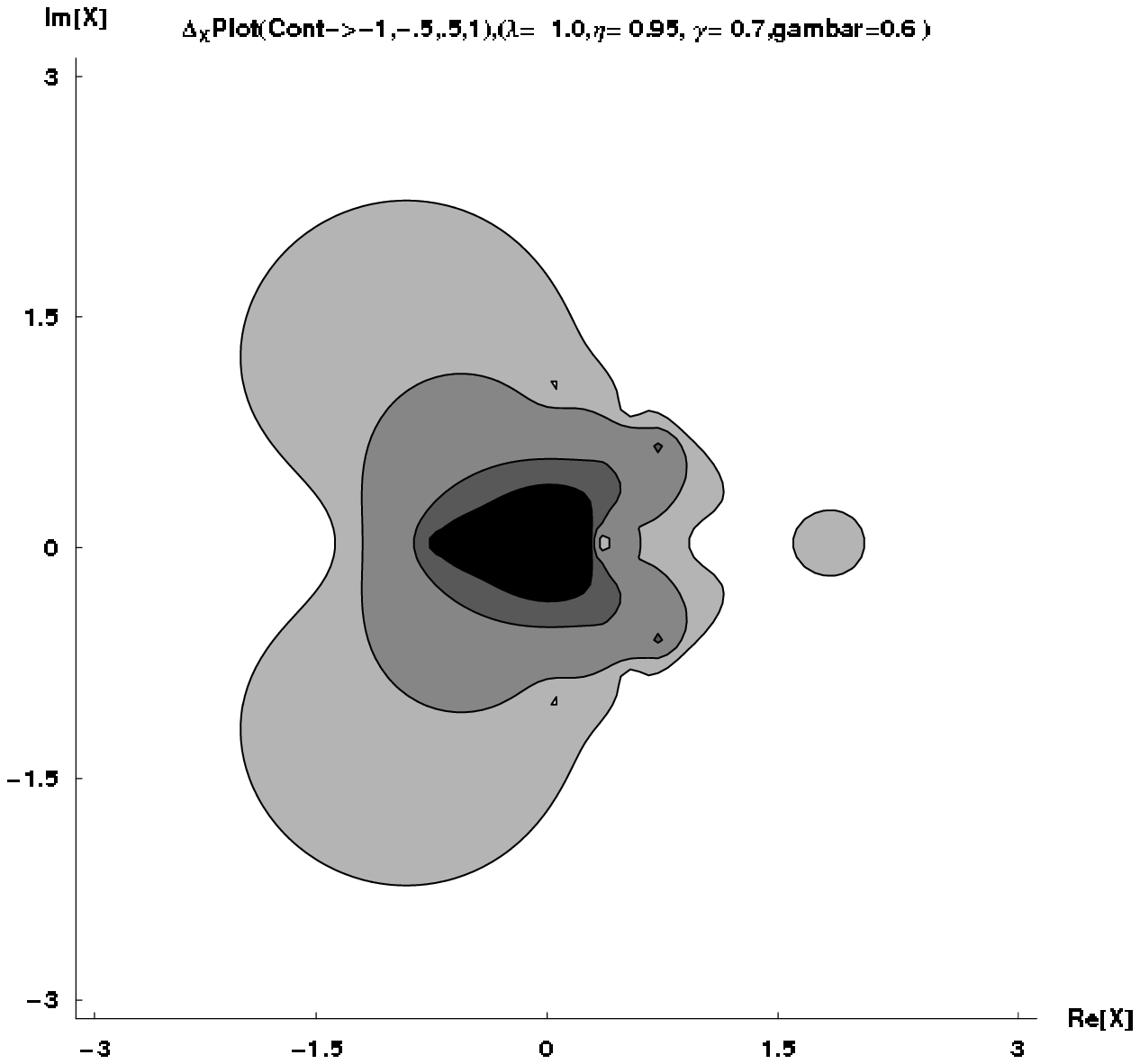}
 \caption{  Contour Plot of the threshold
corrections $\Delta_X$ to $Log_{10} (M_X/1 GeV ) $  on the complex
$x$ plane. Contours at $\Delta_X  = -1,-.5,.5,1$.  The shading
progresses from black ($<-1$) to white $>1$. }
\end{center}
\end{figure}

\begin{figure}[h!]
\begin{center}
\epsfxsize15cm\epsffile{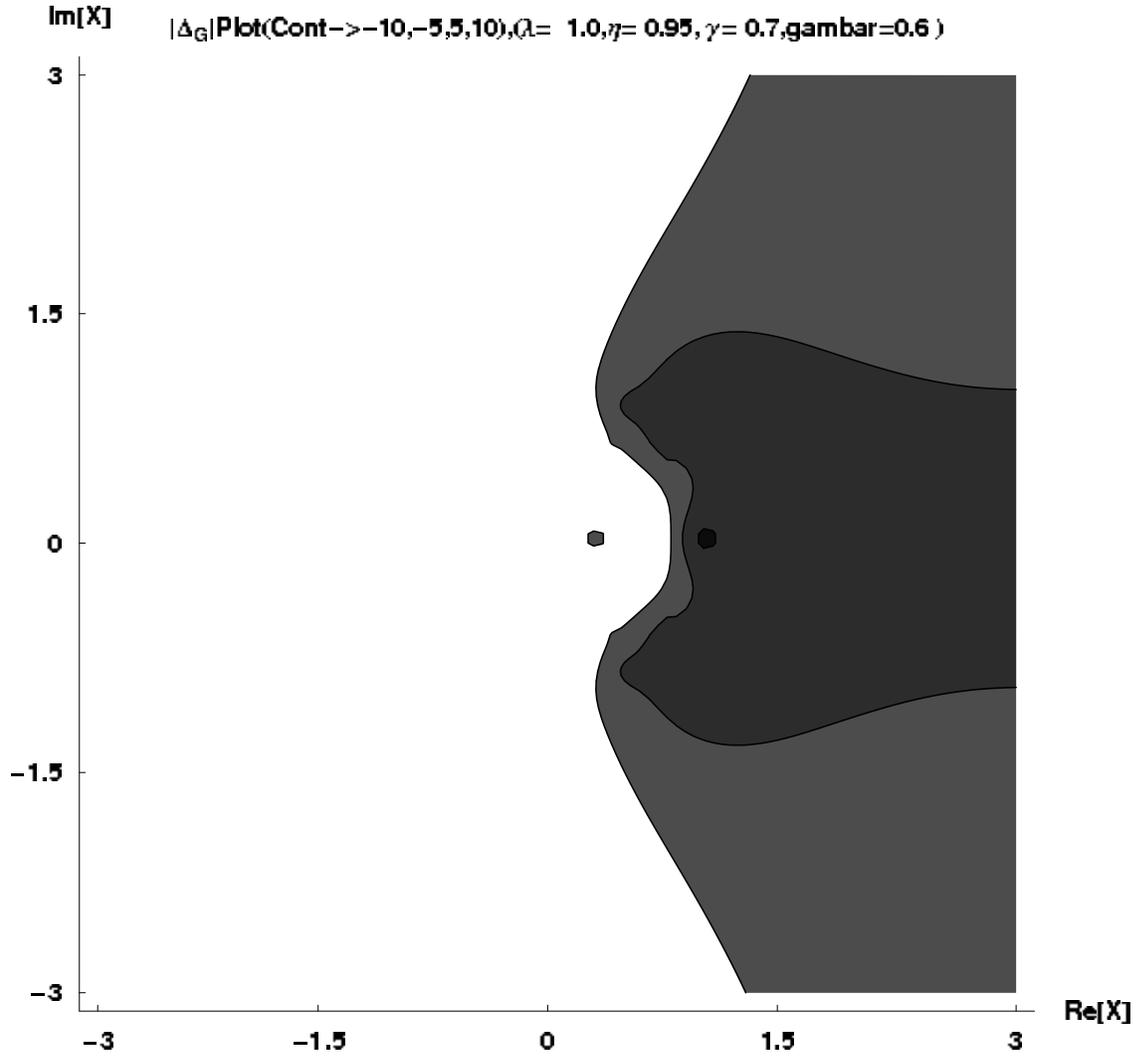}
 \caption{  Contour Plot of the threshold
corrections $\Delta_G$ to $\alpha_G^{-1}(M_X) $  on the complex
$x$ plane. Contours at $\Delta_G = -10,- 5, 5,10$. The shading
progresses from black ($<-10$) to white $>10$. }
\end{center}
\end{figure}

\begin{figure}[h!]
\begin{center}
\epsfxsize15cm\epsffile{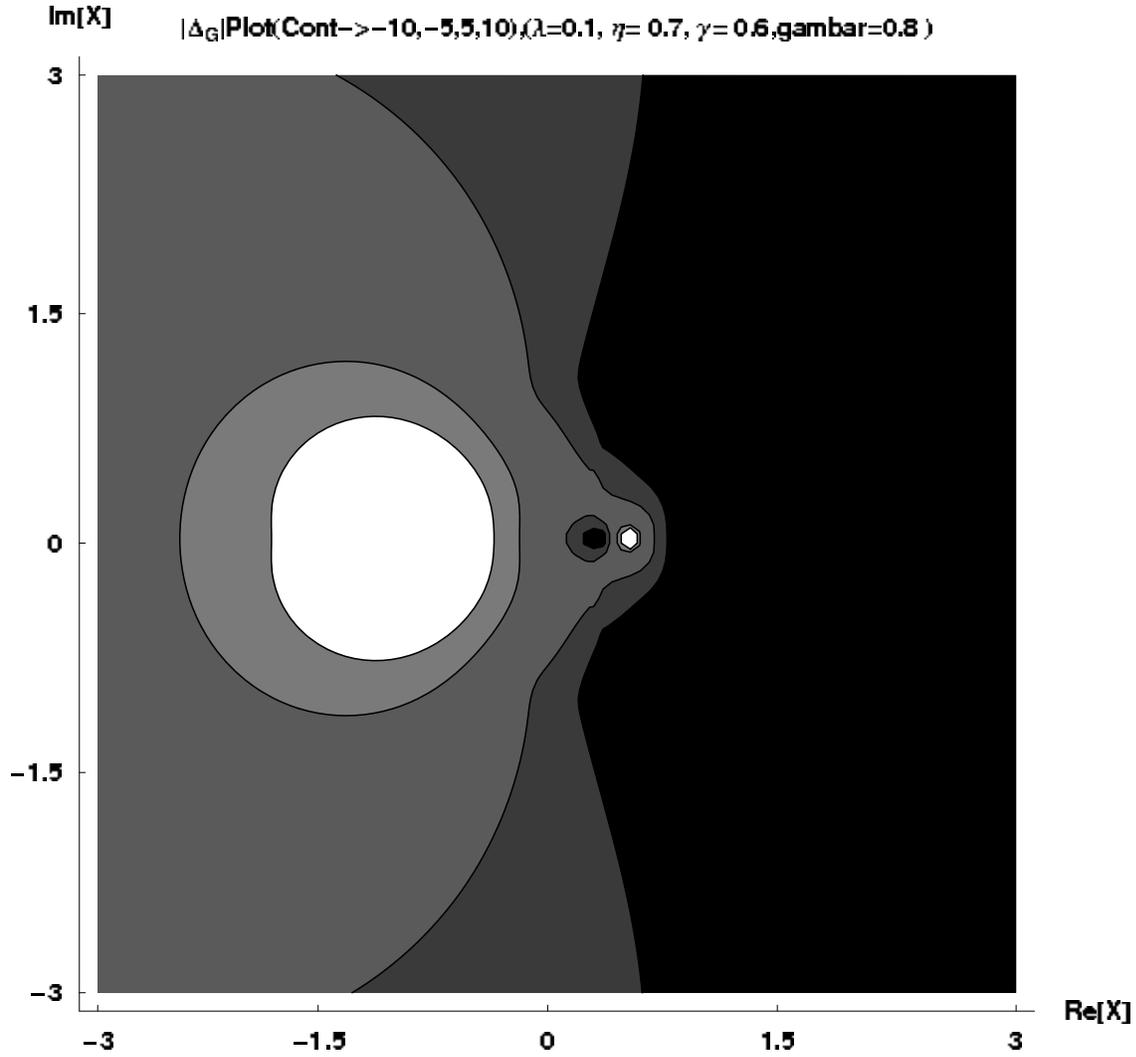}
 \caption{  Contour Plot of the threshold
corrections to $\alpha_G^{-1}(M_X) $  on the complex $x$ plane
illustrating the drastic effect of decreasing the value of the
diagonal parameter $\lambda$. Contours at $\Delta_G = -10,- 5,
5,10$ the shading progresses from black ($<-10$) to white $>10$. }
\end{center}
\end{figure}

\begin{figure}[h!]
\begin{center}
\epsfxsize15cm\epsffile{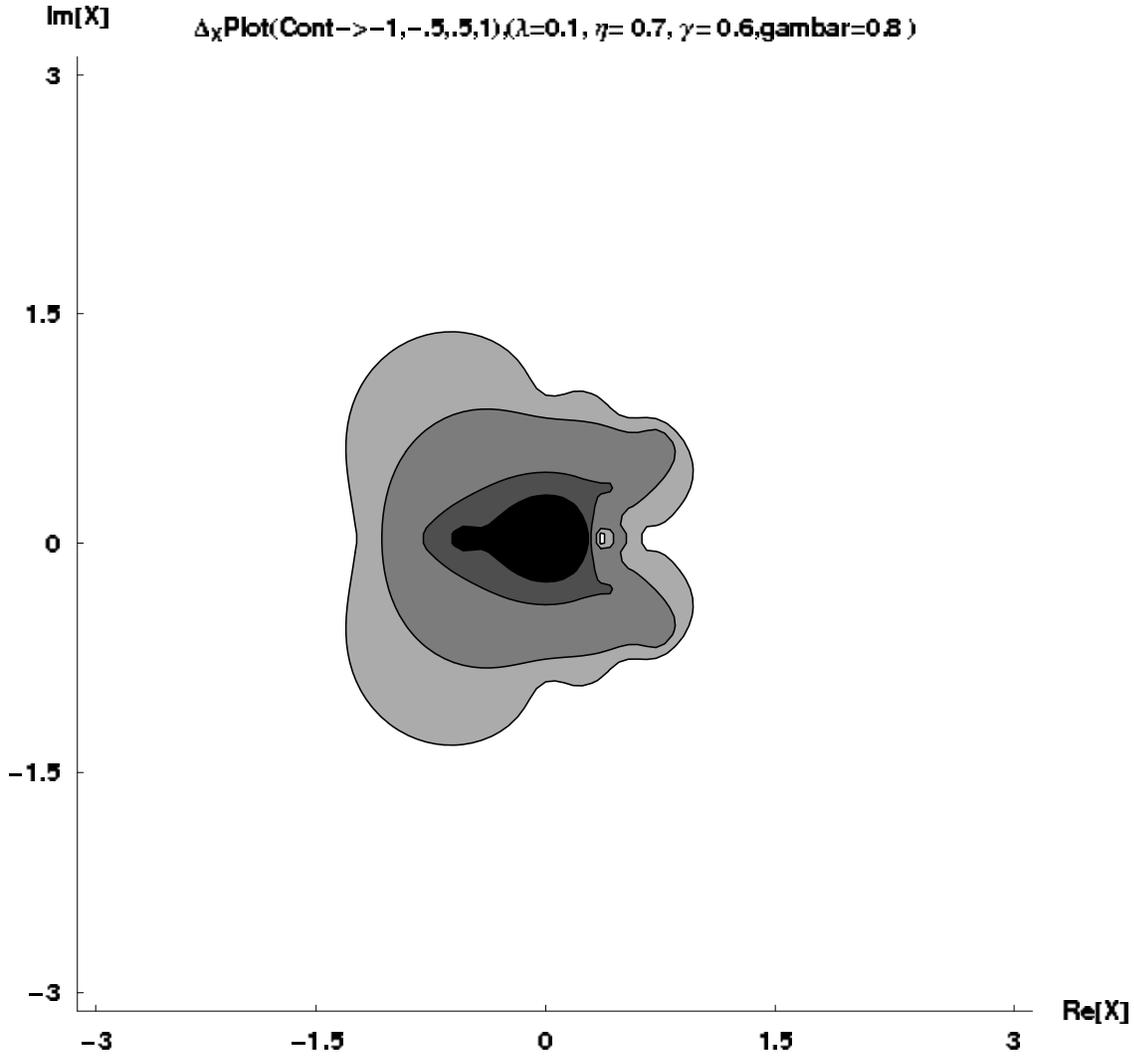}
 \caption{  Contour Plot of the threshold
corrections $\Delta_X $ to $Log_{10} (M_X/1 GeV ) $  on the
complex $x$ plane showing the minor effect of reducing the value
of the diagonal parameter $\lambda$. Contours at $\Delta_X =
-1,-.5,.5,1$ the shading progresses from black ($<-1$) to white.
$>1$ }
\end{center}
\end{figure}

\begin{figure}[h!]
\begin{center}
\epsfxsize15cm\epsffile{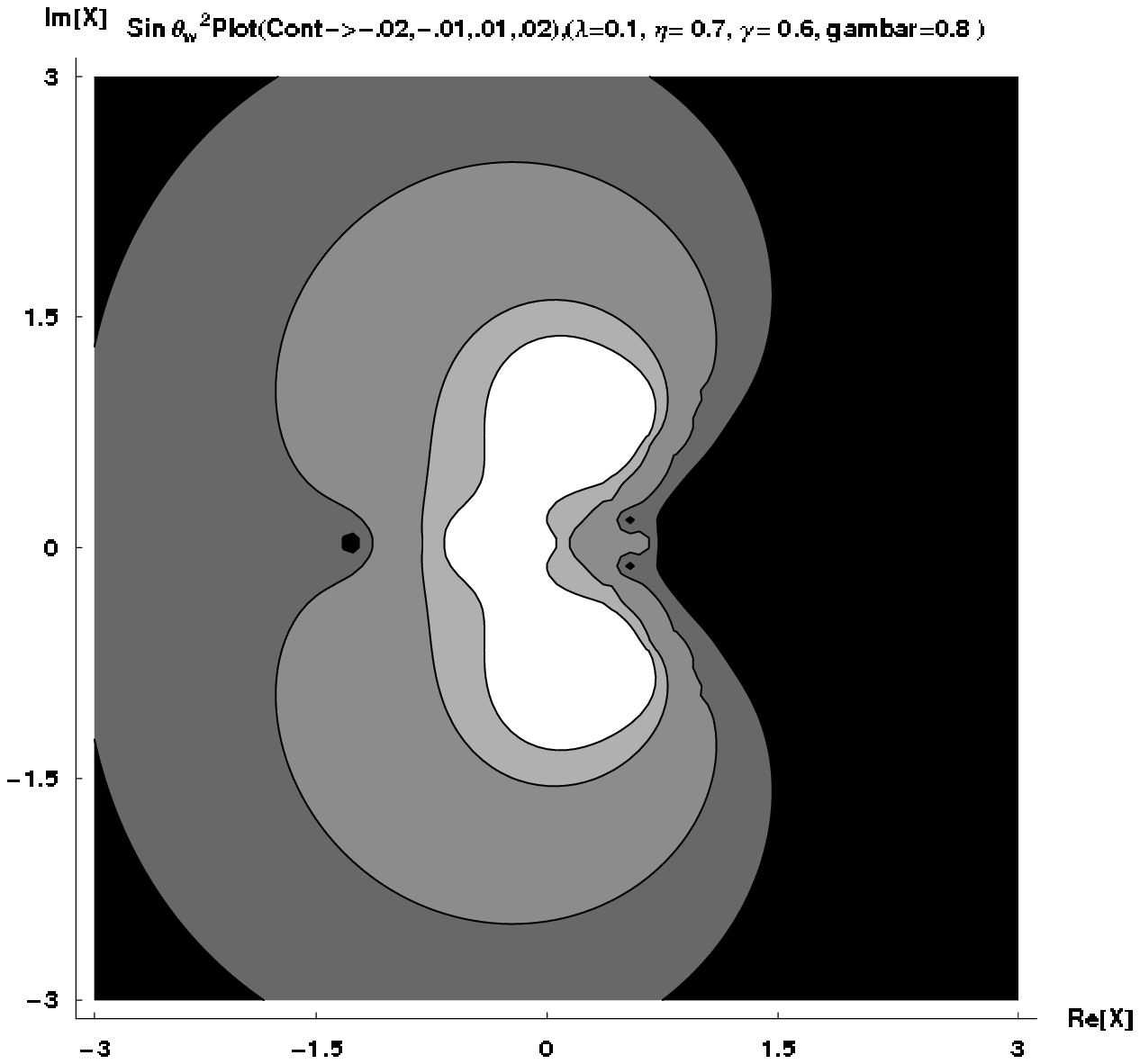}
 \caption{  Contour Plot of the threshold
corrections to $Sin^2\theta_w$  on the complex $x$ plane showing
the minor effect of reducing the value of the diagonal parameter
$\lambda$. Contours at $\Delta_W =-.02,-.01,.01,.02$  the shading
progresses from black ($<-.02$) to white $>.02$. }
\end{center}
\end{figure}

 The parameter  ratio   $m/\lambda$ can be extracted
 as the overall scale of the vevs.   Since the threshold
 corrections    we calculate are dependent  only on (logarithms of)
  ratios of masses,   the parameter $m$  is  fixed   in terms of the
  mass $M_V=M_X$   of the lightest superheavy vector
   particles mediating   proton decay by :
\bea \Delta_X &=& \Delta (Log_{10}{{M_X} \over {1 GeV}})\nonumber \\
  | m| &=& 10^{16.25 + \Delta_X }
   {{|\lambda|}\over
   {g\sqrt{ 4 |\tilde{a} + \tilde{w}|^2 +
   2 |\tilde{p}+ \tilde{\omega}|^2 }}} GeV \label{mvalue}\eea
The presence of the factor $10^{\Delta_X}$ in the formulae for the
neutrino masses will play a crucial   role as one of the hands of
the ``scissor''   operated by Baryon Decay and  constraints and
neutrino oscillation   data.
  The  lowering of the ``slow diagonal''   parameters
  $\lambda,\eta$   tends to make fields light and thus
  give large  negative   corrections to
  $\Delta_G $ (as can be seen in the expansion of the darker areas
   in Fig.4 relative to Fig.3). Such growth of the gauge coupling can
     invalidate the self consistency  of
  the  perturbative  approximation used through out.
  Conversely when $\alpha_G $ decreases too much the
   neglect of Yukawa loops becomes moot. Thus $\lambda,\eta$   are effectively
  restricted  to  magnitudes    of order 1.
     Note, however, that $\Delta_W,\Delta_X$ are insensitive to
   the precise    values -  whether real or complex -  of $\lambda,\eta$
   as long as their magnitudes are  $\sim 1$ : as can be
    seen in the pairs  Fig.1 and Fig. 6  and Fig. 2 and Fig. 5.
    In addition there are the ``slow off diagional''
   parameters $\gamma , \bar\gamma$ whose  effect is quite  mild,
   and can be taken to be any values $\sim$.1 to 1., without any
   appreciable change in the numerical plots.

\subsection{Fermion mass formulae and SMPs}

The Dirac masses of matter fermions in the MSGUT at scale $M_X$
are \bea
 m^u &=&  v( {\hat h} + {\hat f})\nnu
m^{\nu}&=&v ({\hat h} -3 {\hat f})\nnu
 m^d &=& { v (r_1} {\hat h} + { r_2} {\hat f})\nnu
  m^l &=&{ v( r_1} {\hat h} - 3 {  r_2} {\hat f})
\label{dirmass} \eea Here we have extracted $v=174 GeV $.
 Note the characteristic (Georgi-Jarlskog)\cite{georjarls}  factors of (-3) in
 the lepton masses
 relative to the quark  masses. Here\cite{ag1,ag2,gmblm}
 \bea
 \hat{h} &=& 2\sqrt{2}h   \alpha_1 \sin\beta
  ;~~~ \hat{f}  = -4\sqrt{\frac{2}{3}} i   f{\alpha_2}
 \sin\beta  \nnu
 r_1 &=& \frac{\bar{\alpha_1}}{\alpha_1}\cot\beta ;~~~
  r_2  =
 \frac{\bar{\alpha_2}}{\alpha_2}\cot\beta
  \eea
 and  ($h,f)$ are the couplings in the MSGUT
superpotential. They  are $3\times 3$ symmetric matrices with rows
and columns indexed by the SO(10) family indices $A,B= 1,2,3$
\cite{abmsv,ag1,ag2,gmblm}. Notice that since $|\alpha_i | \leq 1$
by unitarity, and $|h,f|< .5$ by perturbativity, the inclusion of
$\alpha_1,\alpha_2$ in the definitions can yield subtleties only
at zeros of $\alpha_1,\alpha_2$.  The Majorana mass parameters
$M_{\nu},M_{\bar{\nu}}$ of the left and right handed neutrinos
(defined as the coefficients of $\nu\nu$, $\bar\nu$ $\bar\nu$ in
the superpotential are\cite{ag2,gmblm} \bea
  M_{\bn} &=&  \hat{f}   \hat{{\bar{\sigma}}}  \nnu
M_{\nu} &=&  r_3 \hat{f}  \label{majmass}\eea
where\cite{ag2,gmblm} \bea \hat{{\bar{\sigma}}} &=& \frac{i
\bar{\sigma}\sqrt{3}}{\alpha_2 \sin \beta} \nnu
M_O &=& 2(M+
\eta(3a-p ))\\ r_3 &=& -2i\sqrt{3}(\alpha_1 \gamma + 2 \sqrt{3}
\eta
\alpha_2)(\frac{\alpha_4}{\alpha_2})(\frac{v}{M_O})\sin\beta\nonumber
 \eea From these one obtains\cite{ag2}
the Type I\cite{seesaw}and Type II seesaw\cite{seesaw2} Majorana
masses (in the conventional normalization $ W=M_{\nu} \nu\nu/2
+...$)of the low energy neutrinos: \bea M_{\nu}^{I}&=&  v r_4
\hat{n} \qquad;\qquad
     M_{\nu}^{II} =  2v r_3 \hat{f}  \nnu
    r_4 &=&{{i\alpha_2 \sin\beta  }
    \over {2 {\sqrt 3}  }} {{  v }\over {\bar \sigma }} \\
    \hat{n}&=& ({\hat h} -3 {\hat f}) {\hat f}^{-1}
    (  {\hat h} -3 {\hat f})\nonumber
     \eea
 For investigation of the Type I seesaw we will find it useful to
 define functions $F_I,F_{II}$ by ($(1.70 \times 10^{-3} eV)=v^2/M_X^0$ and
    we have  omitted irrelevant phases)
 \bea
M_{\nu}^I &=& (1.70 \times 10^{-3} eV) ~ {{\sin \beta} F_{I}}~
\hat{n}\nnu
 M_{\nu}^{II} &=& (1.70 \times 10^{-3} eV) ~{{\sin \beta} F_{II}}
 ~\hat{f}
 \eea where
 \bea
 \hat{\sigma}&=& \sqrt{\frac{x(1-3x)(1+x^2)}{(1-x)^2}}\nnu
 F_I &=& 10^{-\Delta_X}{\frac {i \alpha_2 } {2\sqrt{6}}}
 (g \sqrt{\frac{\eta}{\lambda}}) \frac{\sqrt{4|\tilde{a}+
 \tilde{\omega}|^2+ 2|\tilde{p}+\tilde{\omega}|^2 }}{\hat{\sigma}}    \nnu
 F_{II} &=&10^{-\Delta_X} ({-2i \sqrt{3}g\alpha_4  }){\frac{(\alpha_1 \gamma + 2\sqrt{3}\alpha_2 \eta)}{\alpha_2 \eta}} \frac{ \sqrt{4|\tilde{a}+
 \tilde{\omega}|^2+ 2|\tilde{p}+\tilde{\omega}|^2 }}{(\xi + 3 \tilde{a}- \tilde{p})}\nnu
\eea
 In these formulae the scale  parameter m has been eliminated
in favour of $\Delta_X$ using eqn(\ref{mvalue}).   Explicit forms
for $F_I,F_{II} $ in terms of the parameter $x$ are given in the
Appendix. We also define the ratio $R=|F_I/F_{II} | $ to capture
the relative strength of the two seesaw contributions. A crucial
point is that if one wishes to maintain perturbativity and work
with GUT Yukawa couplings $\leq 1$ then the large $\tan \beta$
scenario is necessary.  The Type I and Type II fits of the fermion
mass-mixing data  are carried out
\cite{japsnu,matsuda,gohmoh,bert,babmacesnu} assuming only the
characteristic form of eqns.(\ref{dirmass},\ref{majmass}) to
derive ``sum rules''\cite{babmoh,japsnu,matsuda,bsv} among the
different mass matrices at the GUT scale. In these generic fits,
the freedom to choose the coefficients $r_1,r_2,r_3,r_4$ is
assumed to be compatible with the theory in which the mechanisms
are embedded. The fits then yield the neutrino masses only up to
an overall scale and the required relative strength of Type I and
Type II is simply assumed. In \cite{gmblm} we showed that these
assumptions are unjustified for the  MSGUT which seems to impose
extreme dominance of Type I over Type II fits and moreover is
incapable of achieving neutrino masses as large as $ .05 eV$as
required by experiment. Our argument was based on the following
observations :
\begin{itemize}
\item When a pure Type II fit is assumed one finds
that\cite{gohmoh,bert,babmacesnu} the maximal value of $ \hat{f}$
eigenvalues is $ \sim 10^{-2}$ while the corresponding values for
$ \hat{h} $ are about $10^2$ times larger. \emph{This is a
structural feature due to the approximate equality of $m_b(M_X)$
and $m_{\tau}(M_X)$ which ensures that no Georgi-Jarlskog type
contribution can be important for the third generation.} As a
result $\hat{n} \sim 10^2 \hat{f}$. This implies that the ratio
$R$ defined above must obey $R \leq 10^{-3}$ for the pure Type II
not to be overwhelmed by the Type I values it implies. Our initial
survey of the MSGUT parameter space showed that, generically,
 this did not happen and
pure Type II could not work. Even near exceptional points no
escape was allowed by the need to maintain viable USMPs. In the
next section we describe a much more exhaustive survey of the
MSGUT to support this conclusion. Note that this is the crux of
the reason for the failure of the Type II seesaw and it applies
equally to   ``mixed Type I- Type II '' fits\cite{babmacesnu}.

\item The Type I fits \cite{matsu0,babmacesnu} yield, typically, $
\hat{n} \sim 5 \hat{f} \sim .3  $ and  thus we see that \emph{the
magnitude of the function $F_I$ will need to be greater than about
$10^2$ } in order that neutrino masses as large as the .05 eV
required for compatibility with neutrino mass squared splittings
as large as $\Delta (m^{\nu})^2_{23} \sim .002 eV^2$ be
achievable. This was  seen to be  generically un-achievable
in\cite{gmblm}. In the next section we prove that this requirement
combined with that of viable USMPs makes it impossible for the
Type I fit to reach viability. This will be interpreted as
indicating a missing piece to the puzzle, the omission of the
\textbf{120}-plet.

The estimate of the  magnitude  of $\hat{n}$ chosen is obviously
critical to the disqualification of the MSGUT Type I masses.
 A value value ${\hat n} \sim 10$ would drastically modify our
conclusions. The 3 generation fits performed so far are all simply
numerical -although guided by analytic insight into where in the
parameter space a fit might occur \cite{ matsu0,babmacesnu}.
Therefore every fit will have its own value of the eigenvalues of
$ {\hat{n}}  {\hat{n}}^\dagger$ which might in principle have
widely different magnitudes. However no such exceptional Type I
fit has been found so far. Although an analytical proof of this is
still elusive - because of the  very  complexity that obstructs
the determination of an analytic  fit, there are heuristic
arguments why a large $\hat n $ fit is unlikely to exist. The
charged fermion fit succeeds because the second generation
couplings of the ${\bf{\oot}}$ dominate those of the $\bf{10}$ so
that the Georgi- Jarskog mechanism can operate.  Although the
opposite ($h_{33} >> f_{33} $) is true for the third generation so
that no such mechanism is effective (as it should not since at
$M_X$ the $b,\tau$ masses are roughly equal) yet the the values of
$\hat f$ in the $2-3$ sector are not so small that the eigenvalues
of $\hat n$ become larger than 1.
 The Type I neutrino masses are found to obey a normal
hierarchy where the second and third generation neutrino masses
are much larger than the first generation so that the $2-3$ block
of $\hat n$   is dominant over the rest of $\hat n$. The large
neutrino mixing in the $2-3$ sector implies that the matrix
elements $\hat n$ in that block are roughly of the same order of
magnitude. These elements are generically found to be of magnitude
less than 1 :  the inverse hierarchical structure of $\hat f$
ensures that the normal hierarchy in $\hat h - 3\hat f$ does not
get amplified by the $\hat f^{-1}$ factor in the Type I seesaw
formula  to the extent of increasing the maximal eigenvalue of
$\hat n$ beyond $.5 $ or so.   An analytic  \emph{proof } of this
is still lacking. Thus in principle there is a loophole which is
still unclosed : exceptional Type I fits with maximal eigenvalue
greater than $10$ , if found, will need to be reanalyzed
separately. We emphasize that the question is not one of finding
improved or optimal fits but only of the order of magnitude of the
maximal eigenvalue of $\hat n$.
  \textbf{\emph{  We shall assume
that such exceptional fits  do not exist. In other words we assume
that the   Type I fits  of \cite{matsu0,babmacesnu}  are generic
and representative as regards the order of magnitude of $\hat
n$.}} It is important    to build a data base or other more
compact description of all passable achieved fits that may be used
to survey all the possibilities.  Thus a naive estimate for the
typical maximal magnitude of $\hat{n}$ : $
{\mathrm{Max}}~{\hat{n}}\sim{\frac{ h_{33}^2}{  f_{33}}}>>1 $ is
actually far off the mark.
 As explained above  the characteristic
off diagonal structure of the successful  Type I  fits in the
generic Babu-Mohapatra proposal, seems to be intimately bound up
with this difference between the naive and actual magnitudes of
(the  eigenvalues of) $\hat{n}$. All these remarks are illustrated
by the values taken from the published example in
\cite{babmacesnu} ( which is, however, not a completely
satisfactory fit since the electron mass is fit only to an
accuracy of about $10 \%$).

\be {\hat{h}}   =  \left(\begin{array}{ccc}
 - .00016  -  .00013   i &    - .000456   -   .00035   i &    - .00642  +
        .002  i  \\  - .00046  - .00035   i & - .00584- 0.00101  &
  0.03142    \\  - .00642
+ .002   i &    .03142   & - .55633  +
         .095531   i\end{array} \right)
 \nonumber \ee

\be {\hat{f}}  = \left( \begin{array}{ccc} 0.00023  + 0.00018 i
& 0.00018  + 0.00017 i & 0.00285  - 0.00053   i \\
   0.00018  + 0.00017   i  & 0.00697 + 0.00164   i
  & -0.01347  - 0.00158   i \\
    0.00285  - 0.00053   i  & -0.01347 - 0.00158   i  & 0.00053 - 0.037858
  i\end{array}\right) \nonumber \ee

\be {\hat{f}}^{-1} =  \left(\matrix{ 2997.76 + 347.393i  & 115.886
-418.543i &  125.607 -173.883i\cr 115.886 -418.543i &
    51.4218 +16.1651i &  -40.8494+15.3566i\cr
    125.607-173.883i &  -40.8494+15.3566 i &  -18.5738+4.48526 i
   \cr  }\right) \ee

\be {\hat{n}}^{est}_{33}=( \hat{h}_{33}-3
{\hat{f}}_{33})^2/{\hat{f}}_{33}
  = 6.37485+6.98329 i\nonumber\ee

 \be {\hat{n}} =  \left(\matrix{  .00278 + .00284  i &   .00712  +  .00287
i &   .02697 + .00393  i\cr   .00712 + .00287 i &
     .09225 + .02520  i &- .12975-0.08165i
    \cr  .02697 + .00393 i & - .12975-0.08165i & 0.00256 +
    0.15193 i  \cr }\right) \nonumber\ee
 \be     ({\mathrm{Eigenvalues}} [ {\hat{\mathbf{n}}}.
      {\hat{\mathbf{n}}}^\dagger ])^{{\frac{1}{2}}}  =
 \{0.2771,0.06298,0.01088\}\nonumber \ee

The magnitude of the maximum eigenvalue of $\hat{n}$ is seen to be
almost 35 times smaller than the ``generic argument'' one! Thus
generic arguments for this magnitude need to be taken with a large
lump of salt !.

\end{itemize}

\section{USMP and SMP plots :  Gloomy Pictures ?}
 We wish to survey the   USMP and SMP
variation over the complex  $x$ plane. We will use generic but
fixed values with magnitudes  $\sim 1 $ for the `diagonal' slow
parameters\cite{ag2,gmblm} $\{\lambda,\eta \} $. We have checked
that for any such (real or complex) values there is no significant
modification of our conclusions.  If one lowers $\lambda,\eta$
below about .2 or so the USMPs exhibit pathologies due to mass
scale lowering. Specifically, an increase of $\Delta_G$ beyond a
value of $10$ or a decrease below (say) $-10$  makes the
perturbative formulae used throughout unreliable. Similarly
varying the `non-diagonal' slow parameters $\gamma,\bar\gamma$ to
small values, though not forbidden by
USMPs\cite{ag2,gmblm} is also infructuous for purposes of viable SMPs.\\
Besides the survey across generic regions of the complex  $x$
 ( equivalently the 3 $\xi $ ) plane(s) we must also
 examine the possibility
  \cite{abmsv,ag2,bmsv,gmblm,bmsv2} that the behaviour at certain
 exceptional points may invalidate the generic trend of seesaw
 exclusion. For this purpose we provide in Table I a list of the
 exceptional points . These consist on the one hand of a list of
 points of enhanced gauge symmetry where the USMPs typically
 show spikes\cite{ag2} and on the other of special points where
 the SMPs may exhibit spikes\cite{gmblm,bmsv2}. The
 complete  analytic\cite{abmsv,bmsv,bmsv2}
 identification of the latter set of  points using the convenient
 parametrization of vevs in terms of Rational functions of the
 variable $x$ alone (i.e with $\xi$ eliminated in favour of $x$)
 is related in the Appendix.
\begin{table}
\begin{center}
\begin{tabular}{|l|l|c|}
\hline $x$        &        $\xi$   &  Symmetry/remarks \\
  \hline
 1/2 &  -5 & $SU(5)$ ,zero of p2\\
 -1 &10 &$SU(5)$   flipped,zero of $p_2$ \\
 0 & 3& $SU(3)\times SU(2)_L\times SU(2)R\times U(1)_{B-L}$\\
1/3 & -2/3 & Flipped $ SU(5) \times U(1)$ \\
$\pm $i & 3$\mp $ 6 i &$SU(3)\times SU(2)_L\times U(1)R\times
U(1)_{B-L}$\\
  0.123765 & 1.93008 & zero of $p_3$\\
 0.646451 - 0.505389 i&4.86831 + 3.29252 i& zero of $p_3$ \\
 0.646451 + 0.505389 i& 4.86831 - 3.29252 i& zero of $p_3$ \\
 $(3 \pm i\sqrt{7})/8$ &    ${3\over 2} (1 \mp  i \sqrt{7})$) &
   zero of $q_2$ \\
-3.46301&28.2958&zero of $p_5$\\
0.262212  $\pm $ 0.388123 i &2.13469   $\mp$ 3.38025 i&zero of $p_5$\\
0.358184  $\pm $ 0.133971 i & -0.449266  $\mp$ 2.30098  i&zero of $p_5$\\
0.1984   & 1.167    &zero of $q_3'$\\
-0.099  $\pm $ 2.24 i & 1.596  $\mp$ 15.572  &zero of $q_3'$\\
\hline
\end{tabular}
\end{center}
\caption{\label{tab2}\em  Special points with  extended gauge
symmetry and/or candidate seesaw enhancement.  }
\end{table}
\subsection{The fatal weakness of Type II }

Let us begin with a discussion of the viability of the Type II
mechanism. The SMP  $R$   we proposed\cite{gmblm}    to monitor
the strength of Type I relative to the Type II seesaw is
particularly simple to analyze. It is   the magnitude of a
rational analytic function ( the ratio of a polynomial of degree
10 to a polynomial of degree 9) of $x$ and is dependent on no
other parameters.

\bea R  = |\frac{F_I}{F_{II}}| =|{\frac{(x-1) (3x-1)q_2
q_3'^2}{4(4 x-1) (x^2+1)q_3^2}}| = |{\frac{P(x)}{Q{(x)}}})|
\label{RX}\eea (the polynomials $q_i,q_i'$are defined in Table II
in the appendix)\\ Therefore  $R$ is bounded from below by $0$ and
its minimae are isolated : they are the zeros of $P(x)$. As
$|x|\rightarrow \infty $ it \emph{grows} as $|x|$ (with
coefficient $ {3}/{64}).$  For $|x|\geq 3 \Rightarrow R\geq
10^{-1} $  or so. This is clearly visible in the 3D plot shown in
 Fig. 7. Thus only for $|x|\leq 3$ do we need to even consider the
possibility of dominant Type II seesaw. This is in fact the region
which contains the zeros of R namely $x=\{1/3,1,\{(3 \pm i
\sqrt{7})/8\}, \{0.198437, -.0992186 \pm 2.24266 i\}\}$ of which
 the latter two sets are the zeros of $q_2$ and $q_3'$
 respectively. A picture of the variation of R in this region
 of the $x$ plane is given by the 3D plots of $R^{-1}$ in
 Fig. 8  which clearly shows that,  as expected,  the zeros of the
 numerator dominate the behaviour.

\begin{figure}[h!]
\begin{center}
\epsfxsize15cm\epsffile{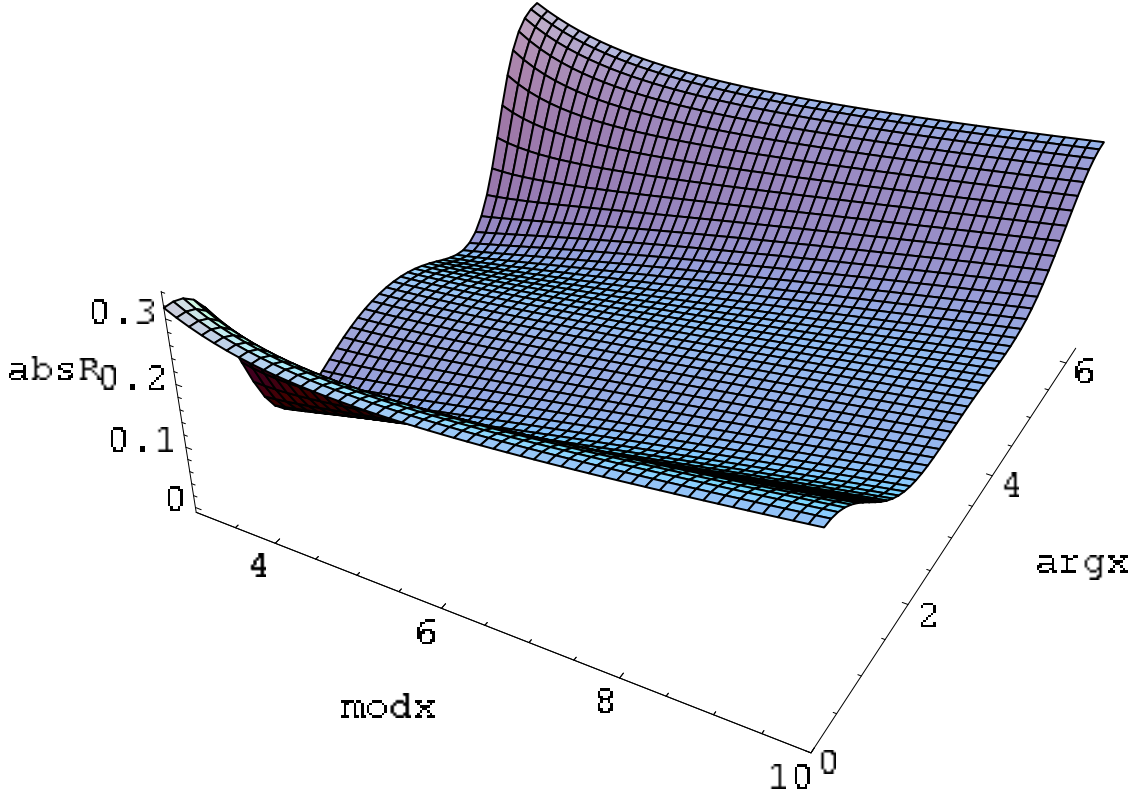}
 \caption{  3D Plot of  the relative strength R
  of Type I and Type II seesaw mechanisms on the complex $x$ plane showing
 that the region $|x|>3$ is irrelevant for promoting Type II dominance.}
\end{center}
\end{figure}

\begin{figure}[h!]
\begin{center}
\epsfxsize15cm\epsffile{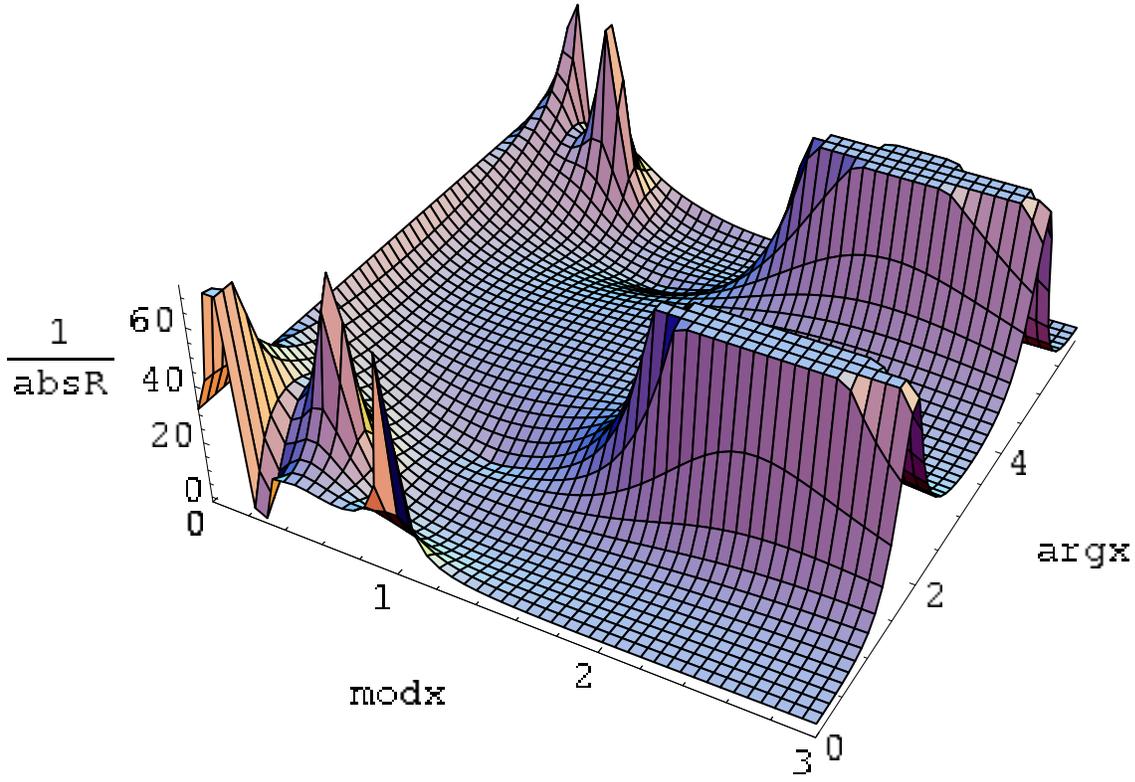}
 \caption{  3D Plot of  the relative strength $ R^{-1} $
  of Type II versus  Type I seesaw mechanisms on the complex $x$ plane showing
  the zeros od $R$ in the  region $|x|>3$.}
\end{center}
\end{figure}

   The real zeros of $R$
  show up as ``cliffs''(instead of narrow peaks )
   due to periodicity in $Arg~ x$ which slices them vertically,
 while  ``ridges'' observed in
 these graphs are associated with the complex zeros of
  $ q_2, q_3'$) (one needs to magnify the
  graph  to see the ridge form of the zeros of $q_2$).
  There is  a strong  preference for a narrow range
   of $Arg~x$ for minimizing R
 (as also seen directly in the plots of R in the neighbourhood
 of these zeros discussed below). Thus the growth of $R^{-1}$
 is seen to be well localized to the regions around the zeros of R.
  From this survey as well as the
 theory of analytic functions it is clear that the behaviour of
  R and the USMPs in the neighbourhood of its zeros will be
   sufficient to decide whether any Type II solutions are viable.
   We now discuss this in greater detail.

 Since Type II dominance  desires \emph{as small} a value
  as possible of R we need to approach as closely
 as viable to the zeros of $P(x)$. Among   the zeros
 of R, however, are also  the zeros of $\alpha_2$ :
\{$x=\{1/3,1\},  \{ 0.1984,-.099\pm 2.24 i\}$  (zeros of
$q_3'$)$\}$. The
  decrease of $|\alpha_2|$ is limited by the requirement
of maintaining perturbativity in the superpotential  Yukawa
coupling $ \sim { {3\hat{f} }/{\alpha_2}}  <  1$.
 Type II dominant fits yield$ \hat{f}$   about $10^{-2}$
\cite{gohmoh,bert,babmacesnu,gmblm}.  Even if we
 are generous with the coefficients ($\sim  3$) that occur
 in the relation,  $\alpha_2 $ \emph{cannot decrease  below
   }$10^{-2}$ : \emph{so we must
 maintain at least  such  distance from  the zeros of $\alpha_2 $}
 so that:  $|\alpha_2| > 10^{-2}$.  The remaining zeros of $R$ are just those of $q_2$.
 We shall see below that approach to these zeros is also
 limited by the  USMPs. Thus R cannot viably decrease below
 $\bar{R}_{min}=Min\{ \bar{R}(x_i) , i=1....7\}$ the values of R at
 closest permissible  approach to its zeros $x_i$.

  No further  survey of the parameter space is thus required
  than to evaluate  $\bar{R}_{min} $.
 To estimate   $\bar{R}_{min}(x_i)$   is
straight forward given our established USMP-SMP numerical
 techniques\cite{ag2,gmblm}. We allow an approach to the zeros of R  as
 closely as will permit $\alpha_2 \geq 10^{-2}$. This limits
 the radius of the  circle of closest approach to the zeros
$x=\{1/3,1,  \{0.198437, -.0992186 \pm  2.24266 i\}\}$ of
$\alpha_2$ ,for the 3 real zeros and the pair of complex zeros
respectively, to be
 $\epsilon_i =10^{-2} \{.7,3.9,4.,.53\}^{-1} =
\{.014,.0025,.0025,.02\} $.  We observe values of
  \bea \{x_i,\epsilon_i,\bar{R}_{min}(x_i),  \{USMPs\}\}
  &=&\{1,.014,.007 ,\{\Delta_G\sim -14,
  \Delta_{W}\sim -.11\}\} \nonumber \\
 &=& \{.3333,.0025, .015 ,\{ USMPs ~ OK \} \}\nonumber \\
  &=& \{.198437,.0025,6\times 10^{-5}  , \{\Delta_X <-2.9 \} \}\\
  &=&\{.198437,.025,  .016 , \{\Delta_X <-2.4 \} \} \nonumber\\
 &=&\{- .099219   +  2.2426    i  ,.02 , 2\times 10^{-5},\nonumber \\
& & \{  USMPs ~ OK, F_{II}\sim
2.4\}\}\label{Rbarvalsal2zeros}\nonumber \eea The closest one
comes to the required values of $R\sim 10^{-4} $ or smaller is in
the case of the complex zeros of $ q_{3}'$. However even in this
case the  value of $F_{II} $ is simply too small
 and yields a maximal neutrino mass of about
  $4 \times 10^{-5} eV$(since $\hat{f}\sim 10^{-2}$).  To
  illustrate the behaviour in this most favourable but still
  unviable case we exhibit the variation of $R,\Delta_X,F_{II}$
  on a circle of radius $.02$ around  a complex zero of $q_3$ as
  Figs. 9,10,11.

\begin{figure}[h!]
\begin{center}
\epsfxsize15cm\epsffile{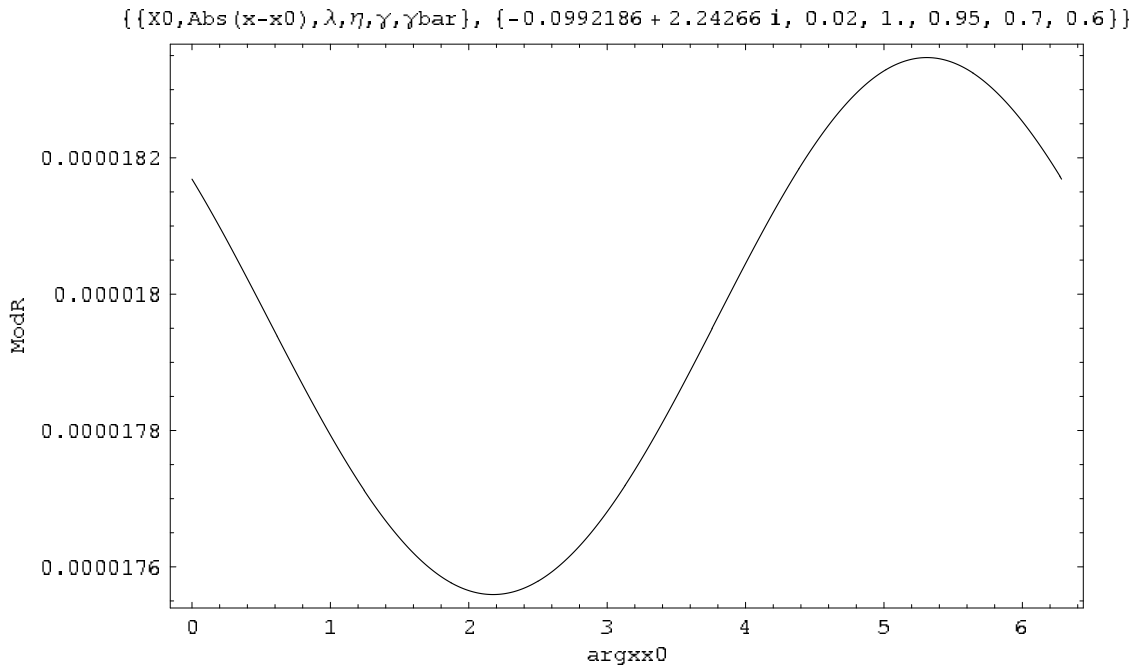}
 \caption{ Plot of $R $
  on a circle of radius $.02$ around  $x_0=-.099219 + 2.2426 i $}
\end{center}
\end{figure}

\begin{figure}[h!]
\begin{center}
\epsfxsize15cm\epsffile{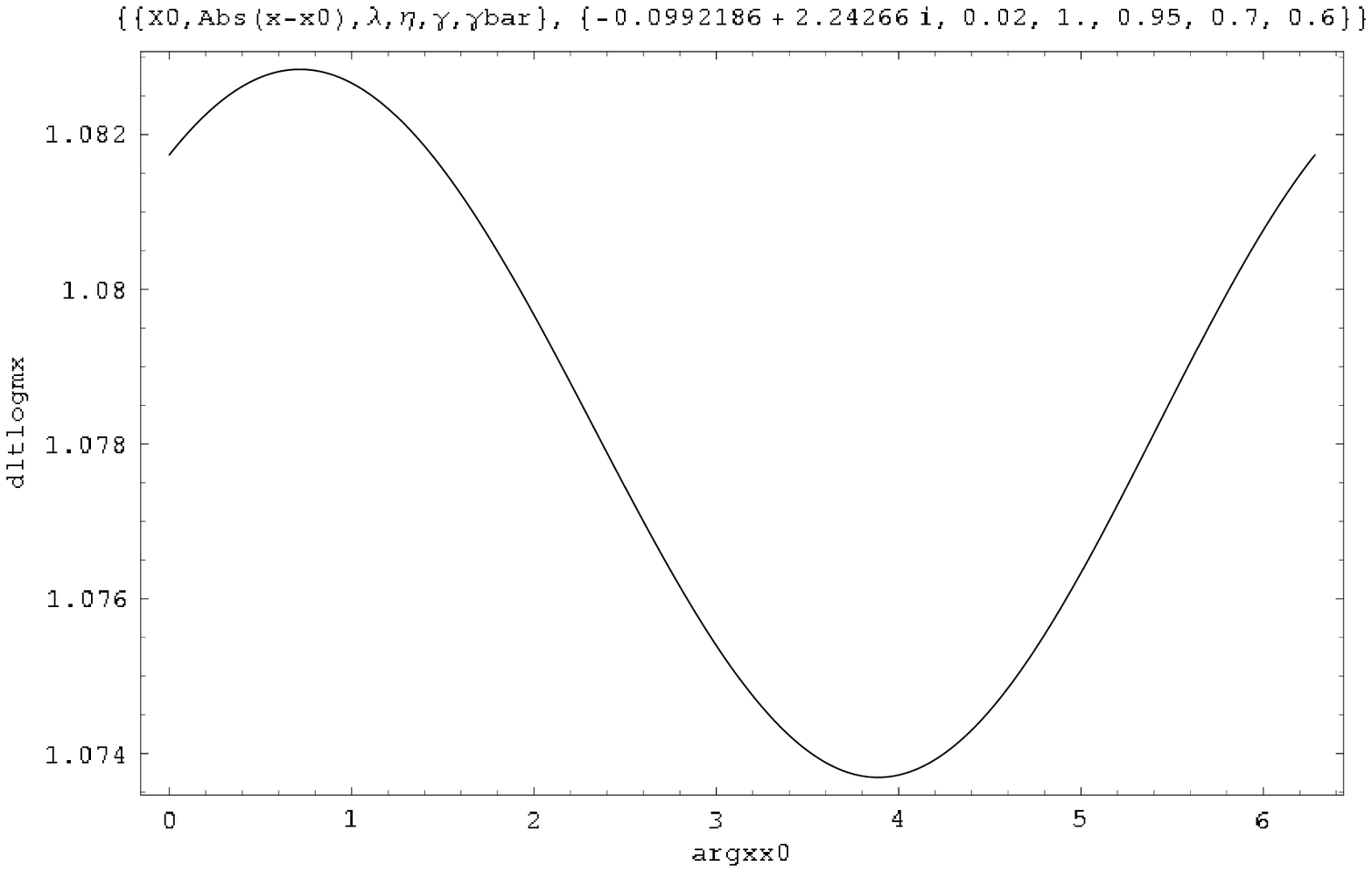}
 \caption{Plot of $ \Delta_X $
  on a circle of radius $.02$ around  $x_0=-.099219 + 2.2426 i $  }
\end{center}
\end{figure}

\begin{figure}[h!]
\begin{center}
\epsfxsize15cm\epsffile{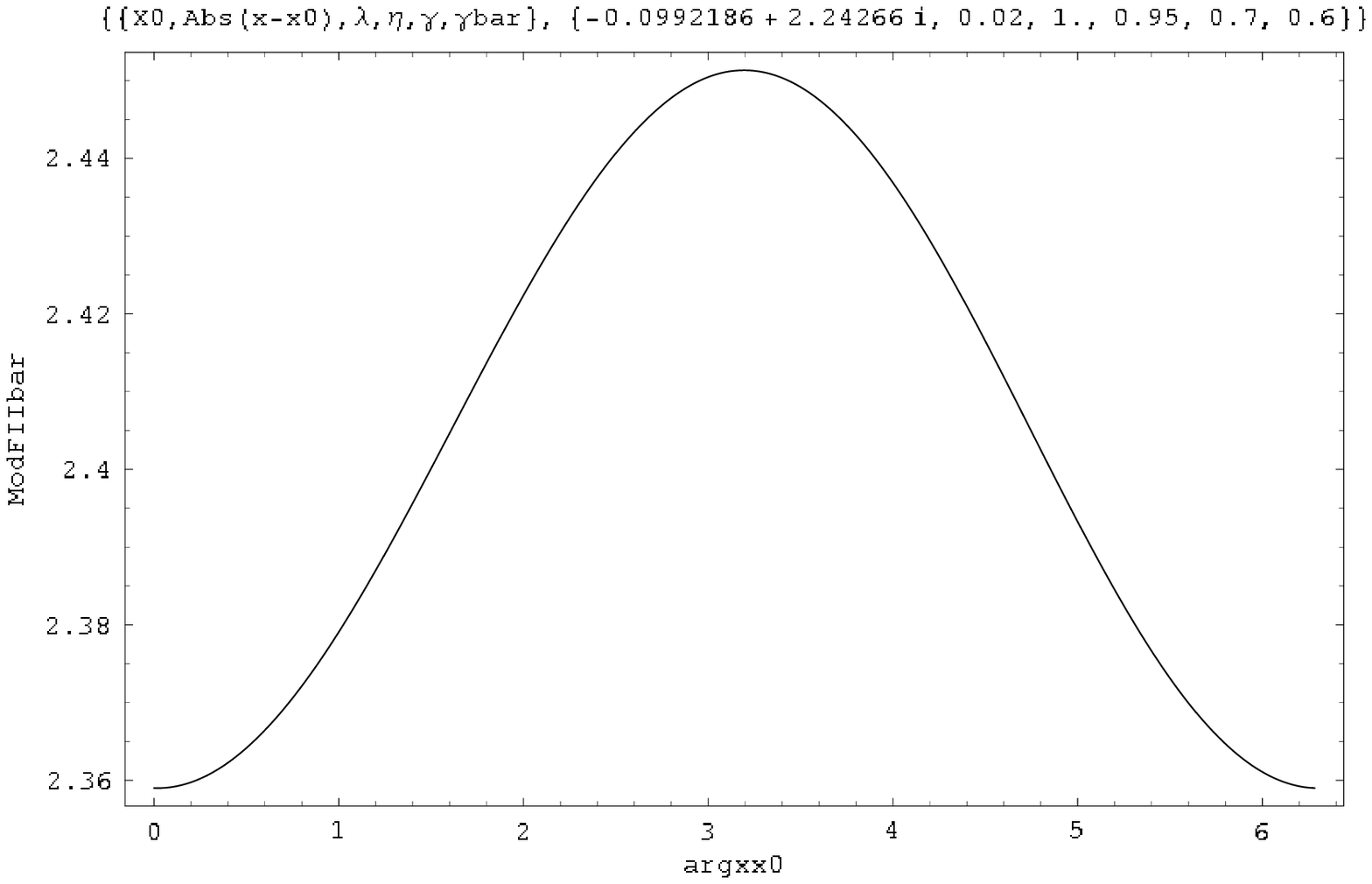}
 \caption{ Plot of $ F_{II}$
  on a circle of radius $.02$ around  $x_0=-.099219 + 2.2426 i $ }
\end{center}
\end{figure}

Finally we turn to the zeros of $q_2$. They are not zeros of
$\alpha_2$ so there is no constraint on approach to the zero from
perturbativity but the conclusions are
 equally negative :

 \bea && \{x,\epsilon,\bar{R}_{min}(x_i),  \{USMPs\}\} = \nnu
 &&\{ (3  + i \sqrt{7})/8 , .06, .015,
 \{\Delta_X <- .9  , F_{II}\sim  80 \}\}
 \label{Rbarvalsal2zeros2}\nonumber\\
 &&\{ (3  + i \sqrt{7})/8 , .005, .0014,
 \{\Delta_X <- 1.3  , F_{II}\sim  2100 \}\} \eea

The rapid fall of $\Delta_X$ (which also promotes $F_{II}$ growth)
as one approaches the zeros excludes them long before the required
small values of $R$ are reached. The strong decrease of $\Delta_X$
for  values of $x$ near $x=0$  is obvious from Fig. 2 and Fig. 5.

To sum up : \textbf{ {the zeros of $R$ are ringed
 by zones of exclusion
 that ensure the parameters of the MSGUT can \emph{never }
be chosen to ensure Type II domination while maintaining viable
USMPs.}}

\subsection{Type I fails : Collapse on the final lap.}

We next turn to a consideration of the overall magnitude of the
Type I seesaw masses which we have shown to dominate the Type II
contribution almost everywhere. The Type I mass formula reads -up
to an irrelevant overall phase-

\bea
 M^\nu_I &=&1.7 \times 10^{-3} eV F_{I}~  \hat{n} \sim\beta \nnu
 F_I&=&
   {\frac{\gamma g}{2\sqrt{2\eta\lambda}}}
 {\sqrt   {\frac{ ((1- 3 x)  }{x (x^2+1)}}}
  {\frac{|p_2p_3| {\sqrt{z_2}}}{\sqrt{z_{16}}}}
 10^{-\Delta_X}q_3'\nnu
  &=& {\hat{F}_I} 10^{-\Delta_X} \label{FI} \eea

The complicated effect  of the RG thresholds and evolution is all
contained inside $\Delta_X$. However this quantity is itself
limited to lie above $-1$ by  Baryon lifetime $>10^{33}$ yrs!
Before actually inspecting the effect of $\Delta_X$ in detail
 we should therefore inspect whether the almost
 completely known function $\hat{F}_I$ could
 ever achieve the  magnitude of 10 or so it would require to
 yield neutrino masses of around $.05$ eV
 (recall $\hat{n}\sim .3$ and $\Delta_X >-1$) in the most favourable case.
This seems a rather mild requirement \textbf{\emph{but we shall
show that it cannot be satisfied viably anywhere in the parameter
space of the MSGUT !}}

The  limiting  behaviour of $|\hat{F}_I|$is

\bea |x| &\rightarrow &\infty \Rightarrow |\hat{F}_I| \rightarrow
\sim.02 \nnu |x| &\rightarrow &  \{0,\pm i \} \Rightarrow
|\hat{F}_I| \rightarrow \infty \label{FIhatlimits}\eea

The approach to the asymptotic value is rather rapid while the
singularities  as $x \sim \{0,\pm i \} $ are quite mild : $\sim
|\Delta x|^{{-}{{1}\over {2}}}$.
   Figs 12,13  which are 3D plots of $|\hat{F}_I|$
  over the complex $x$ plane for $|x| <10$ (more than sufficient to
  reach the limiting form) and $|x|<1.5$ (to show the
  singularities at $x=0,\pm i$ more clearly).

\begin{figure}[h!]
\begin{center}
\epsfxsize15cm\epsffile{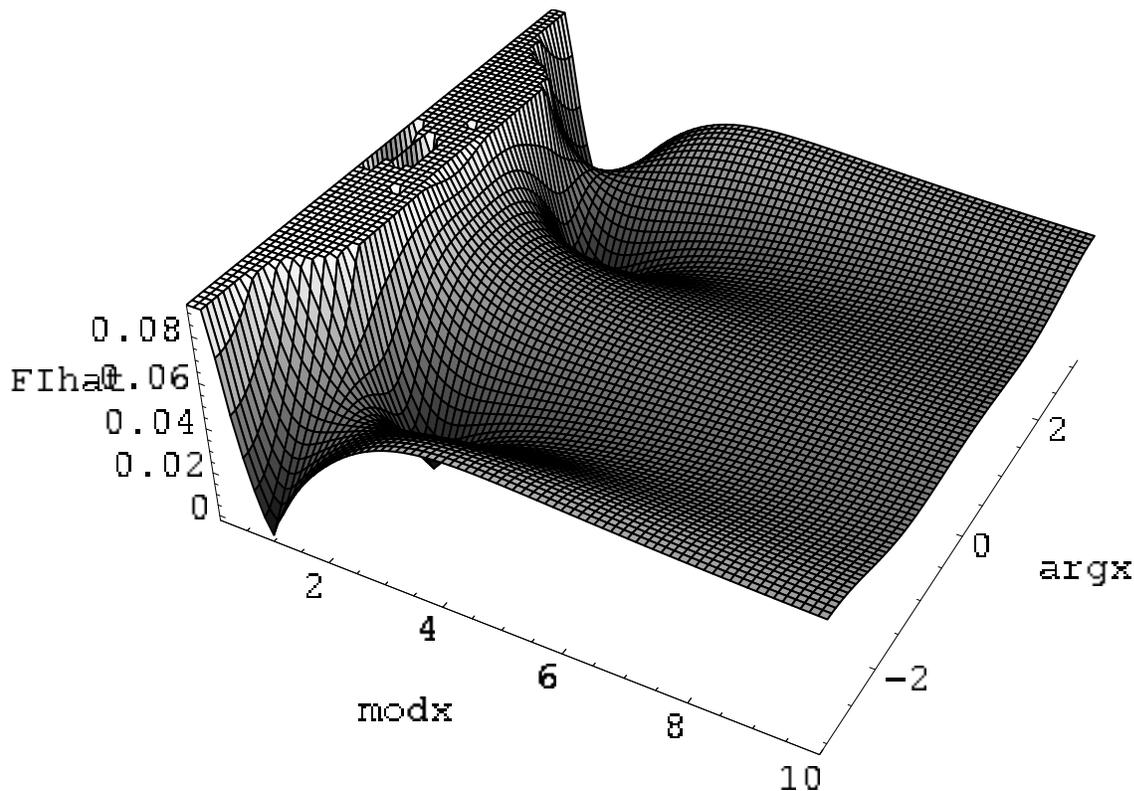}
 \caption{ 3 D Plot of $ |{\hat F}_{I}|$
   on the $x$-plane. }
\end{center}
\end{figure}

\begin{figure}[h!]
\begin{center}
\epsfxsize15cm\epsffile{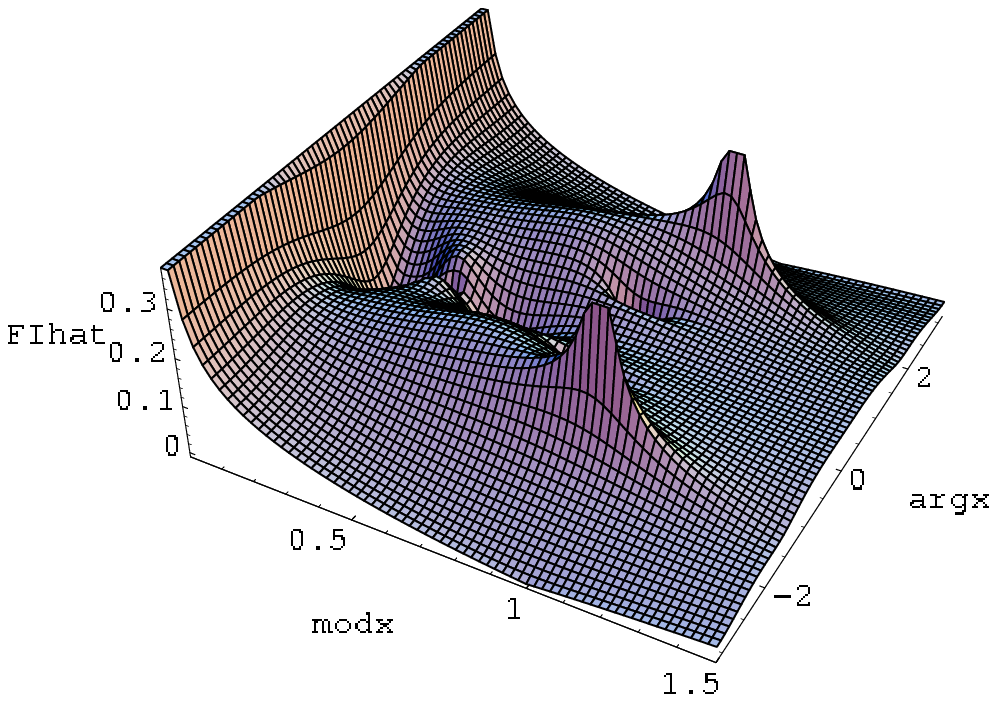}
 \caption{ 3 D Plot of $ |{\hat F}_{I}|$
   on the $x$-plane for $|x|<1.5$. }
\end{center}
\end{figure}

   The dependence of $\hat{F}_I$ on
 the `slow AM'   parameters
 $\{\gamma,\bar{\gamma},\eta,\lambda\}$ is rather
  marginal- as long as we do not consider  pathological cases
  where $  \eta$ or $\lambda $ are very small (
  forbidden by the USMP variation \cite{gmblm}) or pathological
   cases where perturbativity in any of
  of these couplings  is lost. We have verified this by examining
  3-dimensional plots of $F_I$ over the complex $x$ plane
  for random complex values of these couplings without finding
  any visible variation.
   Apart from the constraint that there magnitudes are greater than $.2 $
    their values can
  be chosen randomly without any appreciable effect.

   In  Fig. 12  and   Fig. 13  the rapid approach to the asymptotic value of
  $|\hat{F}_I|$  beyond $|x|=1.2$, the ridge associated
  with $x=0$ (due to its indeterminate phase),  and the pointed
  ridges associated with $x=\pm i$, as well as the boundedness by
  about $.1$ except right close to the singularities, are the
  relevant and crucial features. Plots  which have very
  different coupling values are almost indistinguishable from
  Figs. 12,13. This is easy to understand in
  terms of the expression for $Z_{16}$ (Table II) since that is where
  the dependence on these   parameters
  enters.     It is evident that the
  variation of the slow parameters has no appreciable effect at
  all. So they can be set   to have any   values with magnitudes $\sim 1$
  without  any loss of generality. The  behaviour of $\hat{F}$ implies
 that only very close to the singular points  $x=\{0,\pm i \}$ is
 there any hope that $|\hat{F}_I|$ will achieve the
 required values $\geq 10$. Elsewhere it is manifestly bounded by
 $.1$ or so and is thus a factor of $10^2$ short of the required
 magnitude(in the most favourable case of $\Delta_X$=-1: which value is itself only
 borderline compatible with Super-Kamiokande data on baryon lifetime )

 In  Fig. 14,15  we  exhibit   contour plots of   ${\hat F}_{I} $ and
 $F_I$respectively   on the complex $x$-plane.
   Evidently only around $x=0,\pm i$ is there any chance of large enough $ {\hat F}_{I}$
However  neither region gives a viable result. Firstly, it is
evident from Fig.2 that the region around $x=0$ is forbidden since
$\Delta_X<-1$. A `scissors' operates between the exploding
 effect of $\Delta_X$ as one approaches $x=0$
 and the rapidly decreasing $|\hat{F}_I|$ as one recedes from it.

   Around $x=\pm i$, $\Delta_X$ is $>1$
and hence will suppress rather than enhance the favourable value
of ${\hat F}_{I} $. This is evident in Fig. 15 where
  $x=\pm i$ are out of the running altogether. One
is again caught  in a scissors . At this point (of enhanced
$G_{3211}$ symmetry)
 $\Delta_X$ is large and \emph{positive } at the singularity. So it
 completely overwhelms the growth of $\hat{F}_I$ due to the
 singularity.

\begin{figure}[h!]
\begin{center}
\epsfxsize15cm\epsffile{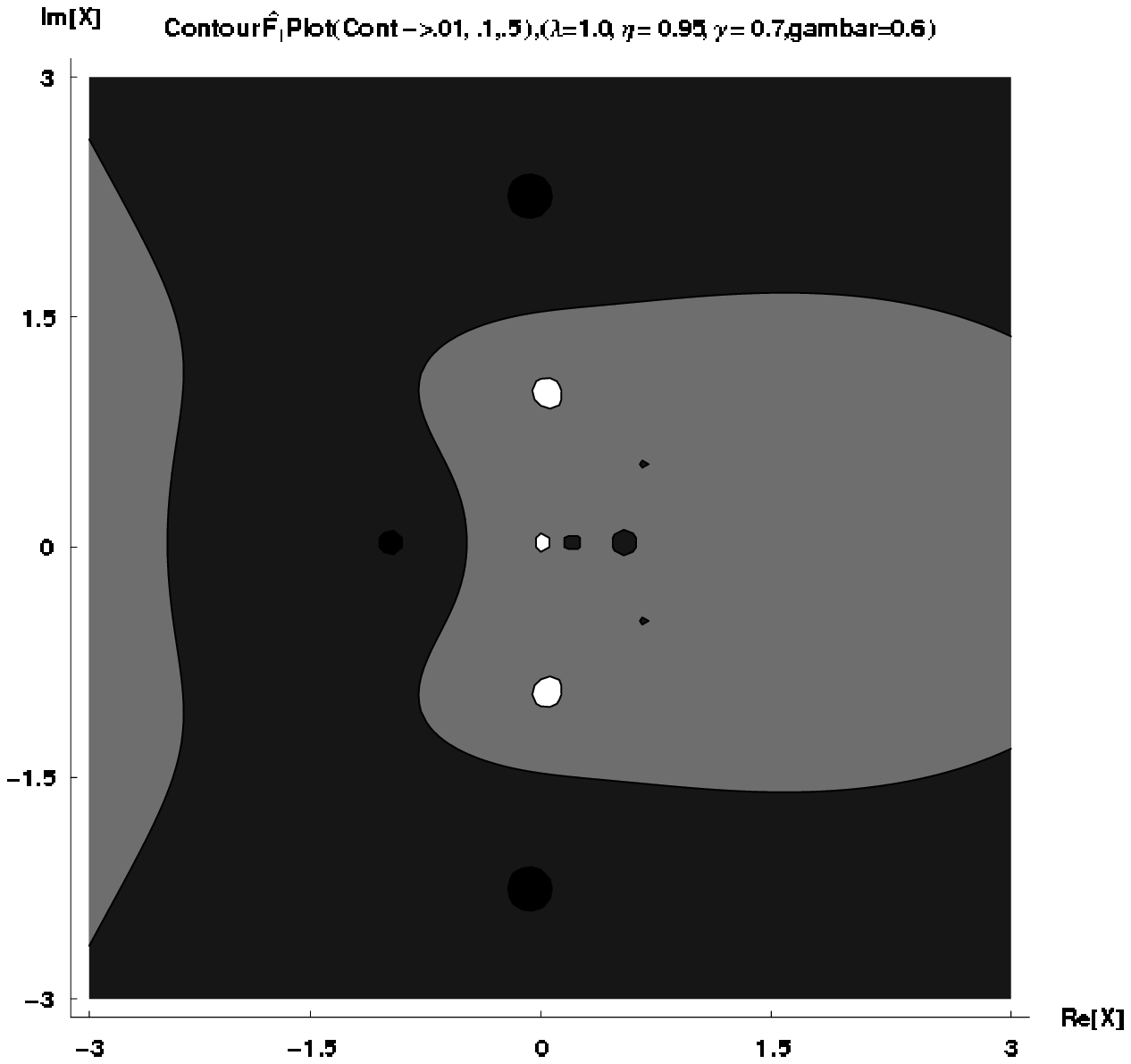}
 \caption{  Contour  Plot of $ |{\hat F}_{I}|$
   on the $x$-plane with contours at
   $ |{\hat F}_{I}|=.1,.9$.  Lower values are shaded darker. The
   small white islands are around $x=0,\pm i$.
     }
\end{center}
\end{figure}

\begin{figure}[h!]
\begin{center}
\epsfxsize15cm\epsffile{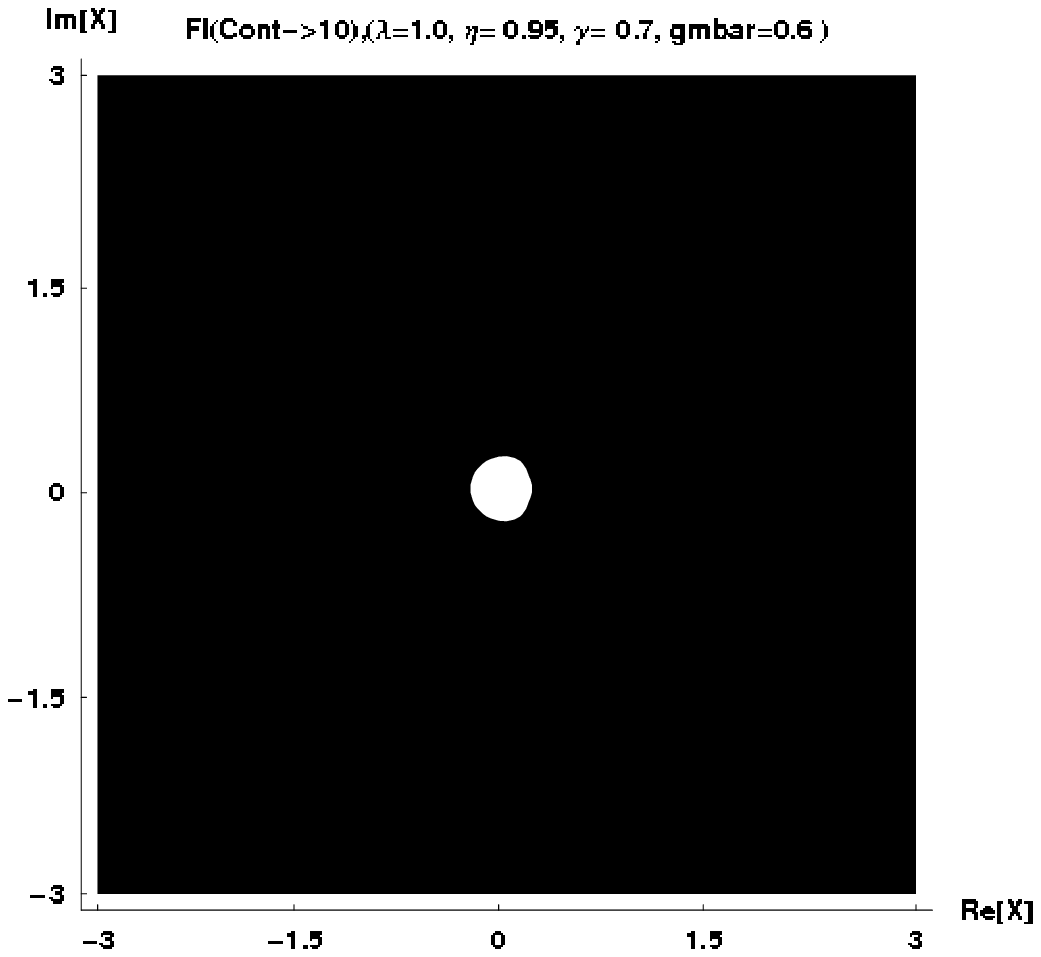}
 \caption{  Contour   Plot of $ |{ F}_{I}|$
   on the $x$-plane with contour  at
   $ |F_{I}|= 10$ lower values are shaded darker. Note that the
   required large values can only lie right close to the forbidden
   region around $x=0$ where $\Delta_X < -1 $.
     }
\end{center}
\end{figure}

     \textbf{\emph{Thus we have
shown that even via Type I seesaw  there is no possibility of
achieving neutrino masses larger than about $5 \times 10^{-3} eV$
in the MSGUT !}}.

Although the argument given above is complete we also studied the
behaviour   at other special  points such as $x=\{-1,1/3,.5\}$ and
the zeros of $\{q_2(x),q_3',p_3(x),p_5(x)\}$, some of which have
been claimed (and some not !) to represent candidate points for
Type I enhancement\cite{bmsv2} but are not in fact so.  We did not
find any appreciable growth of $F_I$ with tolerable values of
$\Delta_X$ at any such point : exactly as expected from our
arguments given above. Some growth is observed for the $x$ values
(digits truncated ) $0.1237$ (real zero of $p_3$), $0.33333$
  (enhanced symmetry point with symmetry $SU(5)_{flipped}\times
U(1))$ and $.198$ (real zero of $q_3'$ ). In all three cases it is
due to catastrophic fall in $\Delta_X$.    The contour plot of
$F_I$ neatly summarizes all this behaviour since only the
forbidden region around $x=0$ ever gives large enough $F_I$.

To trace the reason for the divergence in our candidates from
those of\cite{bmsv2} is easy : if we translate the expressions of
\cite{bmsv2} which are written  in terms of $m^l,m^d-m^l$  into
our expressions written in terms of $\hat{h},\hat{f}$ we
immediately obtain that the functions $f_I,f_{II} $ of
\cite{bmsv2} need to be multiplied by (in our notation)
$\{r_1^2/(4r_2) , r_2 \}\sim \{(p_3 p_4)/(p_5 p_2), (p_5 p_2)/(p_3
p_4) \}$ respectively and this then removes the aforementioned
points from the list of candidates , leaving , after all repeated
factors are cancelled, only $\{0,\pm i\}$ as candidates for strong
Type I seesaw  and the zeros of R namely $x=\{1/3,1 \}$ and  the
zeros of $q_2$ and $q_3'$.

 \section{The Next  MSGUT}
 We have seen that the Type II seesaw mechanism does not work when
 only the $\bf{10}$ and $\bf{\overline{126}}$ FM Higgs irreps are
 employed and even the Type I mechanism fails even though not as
 badly as Type II. It is natural then to look for a role for the
 remaining FM Higgs multiplet type allowed by SO(10): the long
 neglected $\bf{120}$-plet irrep. This irrep has a Yukawa coupling
 matrix that obeys $g_{AB}= -g_{BA}$ i.e. is family index
 antisymmetric. As such it introduces novel properties and
 problems into the fitting matrix problem. It has so far been cast
  in  a minor supporting role\cite{moh120,bert} in the story of the seesaw.
 Only a few authors\cite{matsuda,oshimo,bs,bmsv3} have considered the
  \textbf{120}plet in a large role and even
 they  did not analyse the fitting problem comprehensively.
  The  $\bf{10}$ and $\bf{120}$ irreps may together successfully explain
 the charged fermion spectrum, particularly since the $\bf{120}$-plet also
 contains,besides a \textbf{(1,2,2)} sub-representation,
 a (\textbf{15,2,2)} subrep
 (w.r.t the Pati-Salam subgroup). Thus it can  also be
 used to implement -albeit in a novel way due to its antisymmetric
 Yukawa couplings- a Georgi-Jarlskog\cite{georjarls}
 mechanism to explain the
 approximate equality of $m_s$ and $m_{\mu}$ at $M_S$ in the MSSM.
 This relieves the $ \bf{\overline{126}} $ of the multiple loads it
 has shouldered so far and  which, finally, the present papers
 show it tends to cast down on the last lap of tests.
 Once the $\bf{\overline{126}}$-plet
 coupling $f_{AB} $ is relieved of the necessity of being
 comparable to the couplings of the $\bf{10}$-plet the Type I
 seesaw mechanism can receive the required further 10-100 fold boost
 in magnitude since the Type I seesaw masses are proportional to
 $ \hat{f}^{-1}$. It should be emphasized that the smallness of $\hat f$ in this scenario
  must be considered as a structurally defining condition rather than as the
   smallness  of a perturbation  because  it leads to light right
   handed neutrinos.
     A similar scenario was considered in\cite{oshimo}
  but analyzed by making various somewhat arbitrary assumptions.
   Thus no  generic analysis  like  that
  for the   ${\bf{10-\overline{126}}}$\cite{bsv,bsv2,gohmoh,bert} system
  is yet\cite{newtas}available to compare  with.

In this way we propose that our results provide a hint of the
direction to proceed: every member of the SO(10) FM Higgs family
will receive its characterstic function, role and gainful
employment. It seems that the Good God may be just at least to the
different breeds  of Higgs particles!\\
To implement our (long felt\cite{monica}) intution of a proper
share and role for  each Higgs type in the intricately wrought
elegance of SO(10) unification we have calculated\cite{newtas} the
complete set of couplings and masses of the Next to Minimal
Supersymmetric GUT or NMSGUT defined by the inclusion of a
{\bf{120}}-plet Higgs into the MSGUT Higgs complement. Here we
restrict ourselves to a brief survey of the essential features.

  In the notation of\cite{ag1} the real vector indices
 of the upper left block embedding of SO(6)
  in SO(10) are denoted by  $a,b=1,2..6$ and of the lower right
block embedding of SO(4) in SO(10) by ${\tilde{\alpha},\tilde\beta=
7,8,9,10}$. The doublet indices of $SU(2)_L$($SU(2)_R$)are denoted
by $ \alpha,\beta =1,2(\dot{\alpha} \dot{\beta}= \dot{1},\dot{2}).$
The index of the fundamental {\bf{4}}-plet of SU(4) is denoted by a
(lower) $\mu,\nu$=1,2,3,4 and its upper-left block SU(3) subgroup
indices are $\bar\mu,\bar\nu=1,2,3 $.

The decomposition of the {\bf{120}}-plet w.r.t $ SU(4)
 \times SU(2)_L \times
 SU(2)_R$ is as follows.

\bea O_{ijk}(120)&=&O_{abc}(10+{\overline{10}},1,1)+O_{ab
\ta}(15,2,2)+O_{a\ta\tb}((6,1,3)+
(6,3,1))+ O_{\ta\tb\tilde\gamma}(1,2,2)\nonumber\\
&=&O^{(s)}_{\mu\nu}({10,1,1})+\overline{O}^{\mu\nu}_{(s)}(\overline{10},1,1)
+{O_{\nu\alpha\dot\alpha}}^{\mu}(15,2,2)\nonumber\\
&+& O^{(a)}_{\mu\nu\dot\alpha\dot\beta}(6,1,3)+
{O^{(a)}_{\mu\nu}}_{\alpha\beta}(6,3,1)+
O_{\alpha\dot\alpha}(1,2,2) \eea (where we have used the
superscripts ${}^{(s),(a)}$ to discriminate the symmetric 10-plet
from the antisymmetric 6-plet). \\
The decomposition of its coupling to 16-plets of SO(10) is
\cite{ag1}
 \bea {1 \over
(3!)}\psi C^{(5)}_{2}\gamma_{i}\gamma_{j}\gamma_{k}\chi O_{ijk}&
=& -2{(\bar{O}^{\mu\nu}_{(s)}\psi_{\mu}^{\alpha}\chi_{\nu\alpha}
+O_{\mu\nu}^{(s)}\widehat\psi^{\mu\dot\alpha}{\widehat\chi_{\dot\alpha}^{\nu}}
)}\nonumber\\
&-&2\sqrt{2}{O^{~\mu\alpha\dot\alpha}_{\nu}}
{{(\widehat\psi_{\dot\alpha}^{\nu}\chi_{\mu\alpha}-
{\psi_{\mu\alpha}\widehat\chi_{\dot\alpha}^\nu})}} \nonumber\\
&-&2({O_{\mu\nu}^{(a)}}^{\dot\alpha\dot\beta}
\widehat\psi^{\mu}_{\dot\alpha}\widehat\chi^{\nu}_{\dot\beta}+\widetilde{O}^
{\mu\nu\alpha\beta}_{(a)}\psi_{\mu\alpha}\chi_{\nu\beta})\nonumber\\
&+&\sqrt{2}O^{\alpha
\dot\alpha}(+\hat\psi_{\dot\alpha}^{\mu}\chi_{\mu\alpha}-\psi_{\mu\alpha}
\widehat\chi_{\dot\alpha}^{\mu}) \eea\\
The $SU(4)\times SU(2)_L \times SU(2)_R $ labels in the
above\cite{ag1} equations are trivially decomposable to SM labels
and thus such expressions -- available for every coupling in the
MSGUT and NMSGUT--constitute a complete solution for the Clebsches
to these MSGUTs.

The 120-plet contributes two pairs $ h_5^{\alpha}=
O_{\dot{1}}^{\alpha}, \bar{h}_5^{\alpha}=
O_{\dot{2}}^{\alpha},h_6^{\alpha}= O_{\dot{1}}^{(15)\alpha},
\bar{h}_6^{\alpha}= O_{\dot{2}}^{(15)\alpha}$\\
of MSSM ([1,2,1]$\bigoplus$ [1,2,-1])doublets. Their Yukawa
couplings to MSSM fermions follow immediately from the above
unitary decompositions:

\bea m^u &=&  v( {\hat h} + {\hat f} + {\hat g} )\quad ;\quad r_1=
\frac{ \bar\alpha_1}{\alpha_1}   \cot\beta \quad;\quad
r_2={\frac{\bar\alpha_2}{\alpha_2}}\cot \beta \nnu
 m_{\nu}&=&v (({\hat h} -3 {\hat f}) + (r_5 -3) {\hat{g}})\quad
;\quad  {r_5}=
\frac{4 i \sqrt{3}{\alpha_5}}{\alpha_6+ i
   \sqrt{3}\alpha_5}
 \nnu
 m^d &=& { v (r_1} {\hat h} + { r_2} {\hat f}  +
r_6 {\hat g}); \quad r_6 = \frac{{{\bar{\alpha}}_6}+ i
\sqrt{3}{{\bar{\alpha}}_5}}{\alpha_6+ i \sqrt{3}\alpha_5}
\cot\beta\nnu
   m^l &=&{ v( r_1} {\hat h} - 3 {  r_2} {\hat f} +r_7 {\hat g});\quad {{\bar{r}}_5}=
\frac{4 i \sqrt{3}{{\bar{\alpha}}_5}}{\alpha_6+ i
   \sqrt{3}\alpha_5}\cot\beta
       \label{120mdir}\nnu
{\hat g} &=&2ig {\sqrt{\frac{2}{3}}}(\alpha_6 + i\sqt
\alpha_5)\sin\beta \quad;\quad  r_7=( {{\bar{r}}_5} -
   3r_6)
 \eea
 which clearly exhibits the
analogy with the Georgi-Jarlskog type couplings in the MSGUT. We
propose  to fit the charged fermion masses using light doublet
components of predominantly ${\bf{10}}$ and ${\bf{120}}$-plet
origin.
   The additional terms in the superpotential do not modify the AM
   SSB and moreover still contain only the  26 MSSM multiplet types that we
   described\cite{ag2}
 for the MSGUT. The corresponding mass matrices- some
   of whose dimensions, like the doublet mass matrix given below,
   are raised relative to the MSGUT- will be given, in our
   notation, and along with the corresponding RG and USMP-SMP
   analysis in\cite{newtas}.  A corresponding cacluation of mass matrices only
   - with very different conventions - is already
   available\cite{fukrebut}.

The extra terms contributed by the {\bf{120}}-plet to the
superpotential are :

\bea W_{120}   &=&\frac{M_0}{2(3!)}O_{ijk}O_{ijk} +
\frac{k}{3!}H_i O_{jkl} \Phi_{ijkl}
 +\frac{\rho}{4!} O_{ijm}O_{klm}\Phi_{ijkl}  \nnu
  &&  \frac{ \zeta}{2(3!)}
+ O_{ijk}\Sigma_{ijlmn} \Phi_{klmn} + \frac{
\bar{\zeta}}{2(3!)}O_{ijk}  {\overline{\Sigma}}_{ijlmn}\Phi_{klmn}
\eea

   The coupling of the P($[3,3,\pm \frac{2}{3}]$)and K($[3,1,\pm
   \frac{8}{3}]$)-type color triplets contained in the 120-plet
   leads to quite novel, family index antisymmetric and even $SU(2)_L$
   triplet mediated,  contributions to the effective LLLL and RRRR
   type $ \Delta B \neq 0$ violating d=5 operators in the low
   energy effective theory\cite{newtas}.
    These may lend additional texture to
   the emerging story of the deep connections between neutrino
   mass and Baryon violation\cite{babpatwil,ag2}.

   The composition of the MSSM $[1,2,\pm 1]$
doublet sector can be deduced from the left and right null
eigenvectors of the $6 \times 6$ mass matrix of the NMSGUT
analogously to those defined for the MSGUT in Section II.\\

{\scriptsize
\[ \left( \begin{array}{cccccc}
-M_H & \bar{\gamma}\sqrt{3}(\omega-a) & -\gamma\sqrt{3}(\omega + a)& -\bar{\gamma}\bar{\sigma}&kp & -\sqrt{3}ik\omega \\
 -\bar{\gamma}\sqrt{3}(\omega+ a)& 0 & -(2M + 4\eta(a+ \omega))&0 & -\sqrt{3}\bar{\zeta}\omega & i(p+2\omega)\bar{\zeta}\\
\gamma\sqrt{3}(\omega-a) & -(2M + 4\eta(a- \omega))&0 & -2\eta \bar{\sigma}\sqrt{3}& \sqrt{3}\zeta\omega& -i(p-2\omega)\zeta\\
-\sigma\gamma & -2\eta\sigma\sqrt{3}&0 & -2m + 6\lambda(\omega-a)& \zeta\sigma & \sqrt{3}i\zeta\sigma\\
pk& \sqrt{3}\bar{\zeta}\omega& -\sqrt{3}\omega\zeta&
\bar{\zeta}\bar{\sigma}& -M_o&
\frac{\rho}{\sqrt{3}}i\omega\\
\sqrt{3}ik\omega&i(p-2\omega)\bar{\zeta}& -i(p+2\omega)\zeta& -\sqrt{3}i\bar{\zeta}\bar{\sigma}& -\frac{\rho}{\sqrt{3}}i\omega& -M_0 - \frac{2\rho}{3}a\\
\end{array}\right)\]}

The left and right null eigenvectors are    calculated after
imposing the light doublet  condition ${Det {\cal H}}=0$.

The relative strength of the contributions, of the ${\bf{10}}$ and
${\bf{120}}$-plets vis a vis the 126-plet, to the charged fermion
mass matrices can be studied by examining the variation of
$\alpha_2/\alpha_{1,5,6}$, $\bar\alpha_2/\bar\alpha_{1,5,6}$ in
exactly the same manner as already done by us in this paper. The
pure form of this scenario can thus be examined directly in terms
of fitting the charged fermion masses and mixings in terms of 10
and 120-plet contributions only. If this succeeds to a within
about 10\%,   the effect of including the 126-plet contributions
can be examined\cite{newtas}. Naturally, the investigation is best
begun with the $2\times 2 $ case as for the $\mathbf{10 +
\overline{126}}$ system \cite{bsv,bsv2,bmsv3}. This  is reported
in  \cite{nmsgutI}.

\section{Conclusions and Outlook}

The Supersymmetric GUT based on the the
${\bf{210\oplus126\oplus\oot}}$ AM  Higgs system is the simplest
Supersymmetric GUT that elegantly (almost, v.supra!)  realizes the
classic program of Grand Unification. Its symmetry breaking
structure is so simple as to permit an explicit analysis of its
mass spectrum at the GUT scale and an evaluation therefrom of the
threshold corrections and mixing matrices relevant to various
important quantities. It implements in a fascinating and elegant
way an intimate (``ouroborotic'') connection between the physics
of Lepton number and Baryon number violation. Since it has the
least number of parameters of any theory that accomplishes as much
this theory merited the name of the minimal supersymmetric GUT or
MSGUT. The same simplicity and analyzability of GUT scale
structure also applies to the theory with an additional
{\bf{120}}-plet, since it contains no standard model singlets, and
thus justifies calling it the Next    to  Minimal (or  New
Minimal) Susy GUT (NMSGUT).
  The small number of Yukawa couplings of the MSGUT makes the fit to the now well
 characterized fermion mass spectra  very tight.
 {\textit{\bf{In\cite{gmblm} we initially observed and in this
  paper we have
 shown that the combined constraints of the seesaw fit and
 the preservation of MSSM one loop gauge unification are
 enough to rule out the MSGUT.}}}

  We  now argue  that the very natural
  inclusion of the third possible
  SO(10)FM Higgs type, i.e.the ${\bf{120}}$-plet, which
  was somewhat arbitrarily\cite{abmrs01}--but very
  fruitfully\cite{babmoh,japsnu,matsu0,matsuda,bsv,bsv2,gohmoh,moh120,bert,babmacesnu}
  --excluded   from  a leading role in the `SO(10)party on the GUT
  scene',
  offers a   natural resolution of the difficulty of the
  MSGUT in achieving
  large enough Type I seesaw masses. In this   scenario the
  {\bf{10}}-plet and {\bf{120}}-plet are primarily
  responsible for the   charged fermion mass fit
   due to a relatively suppressed
  contribution of the $\overline{\bf{126}}$-plet to the  charged fermion
  masses due to a small value of the Yukawa coupling
  ${\bf{16\cdot16\cdot\oot}}$. This very suppression implies that
  the Type I seesaw will become enhanced essentially due to
  lowering of the righthanded neutrino masses. Thus far from being
  a perturbation the role of the $\bf\oot$ is rather to define a
  very specific class of Grand Desert in which the right handed
  neutrinos have masses considerably less than the generically
  expected GUT  scale masses. Thus this proposal, which was thrust upon us by detailed
  analysis of compatibility between GUT and Neutrino mass scale hierarchies,
    is distinct from  previous scenarios and, moreover,  has very
  clear and distinct phenomenological consequences.
   Furthermore the $\overline{\bf{126}}$-plet
  Yukawa coupling is itself released from the Type I charged
 fermion constraints and is subject only to the relatively mild
  constraints coming from the right handed Majorana neutrino mass
  limits from cosmology. If this does not conflict with the \
  requirement to achieve the small-large /quark-lepton
   mixing duality then   the one degree of
  magnitude failure of the seesaw in the MSGUT should be quite
  surmountable. In a slogan:

  $\\$

  \begin{large}
   {{\bf{\emph{The MSGUT is dead! Long live the (Next)MSGUT!}}}}
\end{large}

\section{Acknowledgments}
 \vspace{ .5 true cm}
 C.S.A is  grateful to  S. Bertolini and M.Malinsky for
providing   a complete set of data  of one of their Fermion Mass
fits and to G.Senjanovic, Alejandra Melfo for correspondance and
encouragement. S.K.G acknowledges financial support from the
University Grants Commission of the Government of India. C.S.A is
very grateful to Sukhdeep Randhawa and Amarjit S. Mundi for
hospitality and help with graphics issues  and to his family,
   Satbir Kaur, Simran K. Aulakh and Noorvir S. Aulakh, for their
patience and good cheer during the very trying completion of this
paper while travelling on vacation.

 {\section{Appendix}} \vskip .5 true cm
 \vspace{ .2 true cm}

In this Appendix we express our formulae\cite{ag2,gmblm} for
neutrino masses etc in terms of the variable x which parameterizes
\cite{abmsv} the fast variation of AM scale  mass matrices in an
elegant and simplifying manner, as has been particularly
emphasized by\cite{bmsv,bmsv2}. This allows the easy analytic
localization of putative points of seesaw growth\cite{bmsv2}. It
enhances confidence in the completeness of the exclusion by survey
\cite{gmblm} of any possible seesaw fit. The use of the variable x
also significantly simplifies the presentation of data since
considering  variation over the x-plane unifies consideration of
the three solutions for $x$ corresponding to a given value of
$\xi$   while retaining complete clarity due to the rational (3 to
one)  map from the $x$-plane to the $\xi$ plane. The process of
translation merely makes explicit in the mass matrices what was
already done\cite{abmsv,ag2,bmsv} when parameterizing the AM vevs
in terms of x. If we substitute for the AM vevs in terms of x, the
equations for the Null eigen-vectors of ${\cal{H}}$ - the doublet
mass matrix of the MSGUT - subject to the condition $Det{\cal{H}}
=0$ become:(note the labelling of doublets  $ h \bigoplus \bar{h}$
=
 $
(H^{\alpha}_{\dot 2},\Sigb^{(15)\alpha}_{\dot 2},
\s^{(15)\alpha}_{\dot2},{{\phi_{44}^{\dot 2\alpha}} \over
\sq})\oplus (H_{\alpha {\dot 1}},\Sigb^{(15)}_{\alpha \dot1},
\s^{(15)}_{\alpha\dot 1}, {{\phi^{44\dot 1}_{\alpha}} \over \sq})
 $)

\bea A&=&\{\alpha_1,\alpha_2,\alpha_3,\alpha_4\}=N
\{\hat{\alpha}_1,\hat{\alpha}_2,\hat{\alpha}_3,\hat{\alpha}_4\}=N\hat
A\nnu \bar{A} &=& \{\bar{\alpha}_1,\bar{\alpha}_2,\bar{\alpha}_3,
\bar{\alpha}_4\}=\overline{N}
\{\hat{\bar{\alpha}}_1,\hat{\bar{\alpha}}_2,\hat{\bar{\alpha}}_3,
\hat{\bar{\alpha}}_4\}={\bar  N}{\hat{\bar A}}\nnu \hat{A}&=& \{
1,\frac{ \left( {\sqrt{3}}\,{ \gamma}\,{p_4}\,\left(  1  - x
\right) \right) }
   {2\,\eta\,{p_5}},\frac{ \left( {\sqrt{3}}\,
   { \bar{\gamma}}\,{p_2}\,
       \left(  1  - x \right)  \right) }{2\,\eta\,{p_3}},
  \frac{{\gamma}\,{q_3}\,  { \tilde{\sigma}}\,\left( x-1   \right) }
   {2\,{\sqrt{\eta\,{ \lambda}}}\,{p_5}}\}\nnu
   \hat{\bar{A}}&=& \{ 1,\frac{ \left(
{\sqrt{3}}\,{ \gamma}\,{p_4}\,\left(  1  - x \right) \right) }
   {2\,\eta\,{p_5}},\frac{ \left( {\sqrt{3}}\,
   { \bar{\gamma}}\,{p_2}\,
       \left(  1  - x \right)
        \right) }{2\,\eta\,{p_3}},
  \frac{{\bar{\gamma}}\,{q_3}\,
  { \tilde{\bar{\sigma}}}\,\left(x -1  \right) }
   {2\,{\sqrt{\eta\,{ \lambda}}}\,{p_5}}\} \nnu
   N&=&\frac{|p_3p_5|}{\sqrt{z_{16}}} \nnu
    \bar{N}&=&\frac{|p_3p_5|}{\sqrt{\bar{z}_{16}}} \eea

The crucial functions $F_I,F_{II},R$ used as SMPs\cite{gmblm} in
our survey now become (up to irrelevant overall parameter phases )

\bea F_I &=& {\frac{10^{-\Delta_X}}{2\sqrt{2}}} {\frac{\gamma
g}{\sqrt{\eta\lambda}}} | p_2p_3p_5| {\sqrt{{\frac{z_2}{z_{16}}}}}
\sqrt{{\frac{(1-3x)}{x(1+x^2)}}}
 {\frac{q_3'}{p_5}}\nnu
F_{II} &=& {{10^{-\Delta_X}}} ~{\frac{2\sqrt{2}\gamma
g}{\sqrt{\eta\lambda}}} {\frac{| p_2p_3p_5|}{(x-1)}}
{\sqrt{{\frac{z_2}{z_{16}}}}} \sqrt{{\frac{( x^2 +1)}{x(1-3x )}}}
 {\frac{(4 x-1) q_3^2}{ q_3' q_2 p_5}}\nnu
R&=& |{\frac{F_I}{F_{II}}}|=|{\frac{(x-1)( 3x-1)q_2 q_3'^2}{8(4
x-1)( x^2 +1)q_3^2}}|
 \eea

 The various polynomials defined in \cite{bmsv,bmsv2} together
 with some additional definitions are given in Table II.

A number of remarks are in order

{$\bullet$}The asymptotic behaviour of \{$\hat{F}_I ,\hat{F}_{II}
,R $ \} as $|x|\longrightarrow \infty$ is \{$ |x|^0,|x|^{-1},|x|
$\} and as $x\longrightarrow 0$ is \{$
|x|^{-{\frac{1}{2}}},|x|^{-{\frac{1}{2}}},|x|^0 $\} so it is
obvious that only a bounded range of (say)
$|x|< 10$ need be considered to check if R is  small or $\hat{F}_I$ is large.  \\\\

 {$\bullet$} For Type II not to be dominated by the Type I
  implies the ratio R should be very small ($ \sim
 10^{-4} $). So it would seem that the enhanced symmetry points x =
0, 1/3 and the additional points x = \{{$ \frac{ 3 \pm i
\sqrt{7}}{8}$}\},\{{0.198437, -0.0992186 $\pm$ 2.24266 i}\}, which
are the zeros of $q_2, q_3'$ may help to strengthen the Type II
seesaw and should be added to our list of exceptional points. If
one extracts  a factor $ \alpha_2^{-2}$ to define
$R'=R\alpha_2^{-2}$ one sees that only the  the zeros of $q_2$ are
of any relevance to stengthening Type II seesaw .   Our different
parametrization of the problem of dominance together with the
extraction of the hidden dependence on the threshold corrections
to $M_X$ which is coded in $\Delta_X$ makes our
analysis\cite{gmblm} different from\cite{bmsv2}. Differences
from\cite{bmsv2} in the points relevant for Type II have emerged:
we exclude $x=0 $ and the zeros of $p_5$,$p_2$ since they cancel
with the normalizations (which we have not left implicit as
in\cite{bmsv}). Moreover we have used expressions in terms of the
actual Yukawa couplings rather than mass eigenvalues which can
conceal compensating growth of Yukawas and coefficients $\alpha_i$
necessitating separate consideration of whether one has ventured
into strong Yukawa coupling regions. {\textit{Nevertheless we have
included all such putative special points points in our survey and
checked that they do not provide the required exceptional
behaviour}}. \\

  {$\bullet$}The functions $z_2,z_{16}$ have no zeros even when
 $\gamma,\bar{\gamma}$ are both zero. When $\gamma,\bar{\gamma}$
  are non zero    the Normalizing factors never diverge.\\

{$\bullet$}The points $x=0, \pm i  $   are the only remaining
candidates to strengthen Type I and are already included in the
list of enhanced symmetry points.

 \begin{table}[t]
\begin{center}
\begin{tabular}{|l||c|}
\hline $q_2 $&$ 4 x^2-3 x+1$\\ \hline $q_3 $&$ 4 x^3-9 x^2+9 x -2
$\\ \hline
$q_3' $&$ p_4/(3x-1)= x^3 + 5x-1$\\
        \hline
  $p_2 $&$ (2 x-1)(x+1)$\\ \hline
$p_3 $&$
12 x^3-17 x^2+10 x-1$\\ \hline $p_4 $&$ (3 x-1)(x^3+5 x-1)$\\
\hline $p_5 $&$ 9 x^5+20 x^4-32 x^3+21 x^2-7 x+1$\\ \hline
$z_2 $&$ 2|x|^2 + |1-x|^2$ \\
        \hline
$z_{16} $&$ |p_3 p_5|^2 + \frac{3}{4}|\frac{\gamma(x-1)p_3
p_4}{\eta}|^2 +  \frac{3}{4}|\frac{\bar{\gamma}(x-1)p_2
p_5}{\eta}|^2 + \frac{1}{2}|\gamma(x-1)p_3 q_3
\sqrt{\frac{x(1-3x)(1+x^2)}{\lambda\eta}}|^2$\\\hline
$\bar{z}_{16} $&$ |p_3 p_5|^2 +
\frac{3}{4}|\frac{\bar\gamma(x-1)p_3 p_4}{\eta}|^2
+\frac{3}{4}|\frac{ {\gamma}(x-1)p_2 p_5}{\eta}|^2 +
\frac{1}{2}|\bar\gamma(x-1)p_3 q_3
\sqrt{\frac{x(1-3x)(1+x^2)}{\lambda\eta}}|^2$\\\hline
$N $&$ {|p_3 p_5|}/{\sqrt{z_{16}}}$ \\
        \hline
$\bar{N} $&$ {|p_3 p_5|}/{\sqrt{\bar{z}_{16}}}$ \\
        \hline
\end{tabular}
\end{center}
\caption{\label{tab1}\em Functions  of $x$ entering the
expressions for the USMPs and SMPs.}
\end{table}

\vspace{ .4 cm}

 {\bf{Note Added :} } Our surmise  was announced in
May/June 2005 (PLANCK05,Trieste, May 2005   and hep-ph/0506291)
and our proof appeared as  hep-ph/0512224v1). In May 2006 a
calculation appeared \cite{brtschmal} confirming  that optimal FM
fits found by the ``downhill simplex'' method  are not viable in
the MSGUT, for exactly the reasons given by us.

\end{document}